\documentclass[useAMS,usenatbib]{mn2e}
\usepackage{natbib}
\usepackage{aas_macros}
\usepackage{graphicx}
\title[C and N in the Galactic disks]{Carbon, nitrogen and $\alpha$-element abundances determine the formation sequence of the Galactic thick and thin disks.}\author[T. Masseron G. Gilmore ]{T. Masseron$^{1}$\thanks{E-mail:
tpm40@ast.cam.ac.uk} and G. Gilmore$^{1}$ \\
$^{1}$Institute of Astronomy, Madingley Road, Cambridge  CB3 0HA, UK\\}

\begin{document}
\date{}

\pagerange{\pageref{firstpage}--\pageref{lastpage}} \pubyear{2002}

\maketitle

\label{firstpage}

\begin{abstract} 
Using the DR12 public release of APOGEE data, we show that the thin and thick disks separate very well in the space defined by [$\alpha$/Fe], [Fe/H] and [C/N].  Thick disk giants have both higher [C/N] and higher [$\alpha$/Fe] than do thin disk stars with similar [Fe/H]. We deduce that the thick disk is composed of lower mass stars than the thin disk. Considering the fact that at a given metallicity there is a one-to-one relation between stellar mass and age, we are then able to infer the chronology of disk formation.  Both the thick and the thin disks - defined by [$\alpha$/Fe] -- converge in their dependence on [C/N] and [C+N/Fe] at [Fe/H]$\approx$-0.7. We conclude that 1) the majority of thick disk stars formed earlier than did the thin disk stars 2) the formation histories of the thin and thick disks diverged early on, even when the [Fe/H] abundances are similar 3) that the star formation rate in the thin disk has been lower than in the thick disk, at all metallicities. Although these general conclusions remain robust,
 we also show that current stellar evolution models cannot reproduce the observed C/N ratios for thick disk stars. Unexpectedly, reduced or inhibited canonical extra-mixing is very common in field stars. While subject to abundance calibration zero point uncertainties, this implies a strong dependence of non-canonical extra-mixing along the red giant branch on the initial composition of the star and in particular on the $\alpha$-elemental abundance.   
\end{abstract}


\section{Introduction}
It is now well-established from analysis of many disk galaxies that Milky Way-like galaxy disks have a two-component structure. A comprehensive set of 21 review articles providing an overview of the development and present status of this knowledge is available in \citet{Gilmore2013}, with a detailed history of the evidence from star count studies in particular available in \citet{Yoshii2013}. Modern studies of the Galactic thick disk began with \citet{Yoshii1982} and \citet{Gilmore1983} utilising photographic star counts.  Later, quantitative high-quality stellar spectroscopic studies showed that the local thin and thick disk have also distinct $\alpha$-element abundance enhancements: {\it cf} \citet{Fuhrmann1998,Reddy2003,Bensby2005,Adibekyan2011,Recio-Blanco2014}.
Suggestions from analysis of the low-resolution SEGUE [$\alpha$/Fe] data that the thick disk was not chemically discrete from the thin disk \citep{Bovy2012}, in disagreement with both other analyses of SEGUE data \citep{Lee2011} have been shown inconsistent with better resolution data in several later studies \citep{Guiglion2015}. We show below the very clear distinction between thin and thick disks manifest in APOGEE data. The several current high-resolution, high-quality large sample stellar spectroscopic surveys underway - Gaia-ESO \citep{Gilmore2012}, APOGEE \citep{Holtzman2015}, GALAH \citep{DeSilva2015} - together with the imminent Gaia data, are producing large precise data sets, to extend our knowledge far beyond either small samples or low-resolution exploratory surveys.  Recent chemo-dynamic models interpret available data in the framework of Galactic formation, with several different scenarios for Galactic disk evolution presented, as for example the study of \citet{Haywood2013}. \citet{Rix2013} provide a helpful recent review. A fundamental challenge in interpretation of the rich available data in terms of a Galactic evolutionary history is the difficulty of assigning ages (relative or absolute) to field stars. Determining stellar ages from observables will always require stellar evolution models, while stellar properties such as mass and luminosity are currently best constrained with asteroseismology and parallax measurements.  While we await Gaia for mapping large samples of stellar populations, samples available for accurate age determination are currently limited to a maximum of a few thousand stars.

Low-mass giant stars ($\rm M<3M_\odot$) are excellent tracers to probe the Galaxy in multiple dimensions (space, kinematics and chemistry) because they are very numerous and they are intrinsically very bright. Therefore they can be observed and studied in detail up to very large distances in the Galaxy (and beyond) with modern instrumentation, including in particular large spectroscopic surveys such as Gaia-ESO and the APOGEE survey. 

The APOGEE survey obtains high quality spectra of Galactic giant stars in the H band, with a reduction system providing homogenously-determined stellar parameters and elemental abundances. The recent public release \citep{Holtzman2015} provides a homogenous set of stellar parameters and elemental abundances, with in particular of relevance for this study, carbon, nitrogen and $\alpha$-element abundances for more than 200,000 stars. While the calibration of abundance and parameter scales derived from automated mass-production procedures is a major study in its own right, with much progress anticipated by comparing the various surveys underway with each other, and with Gaia and asteroseismology results over the next few years, already with appropriate care this large dataset provides a powerful opportunity to study Galactic star formation histories, and both stellar and Galactic evolution.   

Of particular interest among the elemental abundances which can be derived from H-band spectroscopy are carbon and nitrogen. The C and N elemental abundances in red giants are sensitive to stellar characteristics including mass and age. Carbon and nitrogen participate in many nucleosynthetic reactions so that their abundances are altered by all stars during their evolution. During some stellar evolutionary phases these altered abundances are dredged up into the stellar atmosphere, providing direct evidence of the stars nucleosynthetic history, which probes stellar age and mass. One must therefore disentangle stellar evolution from Galactic evolution, but carbon and nitrogen provide the information to constrain both star formation and galactic evolutionary histories.   This analysis can be further extended by utilising the $\alpha$-element abundance, which is unchanged during stellar evolution, and is a known indicator of star-formation histories, and distinguishes the thick and the thin Galactic disks.  

In the present paper, we take advantage of the measurements of C, N and the $\alpha$-elements in the large sample of giants in the APOGEE survey and consider their distribution in the context of state-of-the art stellar evolution models.

\section[]{Data selection and stellar evolutionary state categorization}
We are interested in the subset of stars which have well-determined relative abundances for carbon, nitrogen, iron, and the alpha elements. In order to ensure optimal data quality, and given the very large APOGEE data set, we are able to sub-select from the full sample only those with signal-to-noise ratio $>$100, while retaining a statistically useful sample size. In \citet{Holtzman2015} the analysis team caution that they still have some challenges in determining accurate parameters for cooler stars, those with $\rm T_{eff} < 4000K$. Therefore we also remove those cooler stars from our discussion: in practise such cool stars are in any case not relevant to this study, as the discussion below explains. This leaves a sample of $\sim$81000 stars having C and N measurements and which met these selection criteria.   

\begin{figure}
\includegraphics[width=6cm,angle=-90]{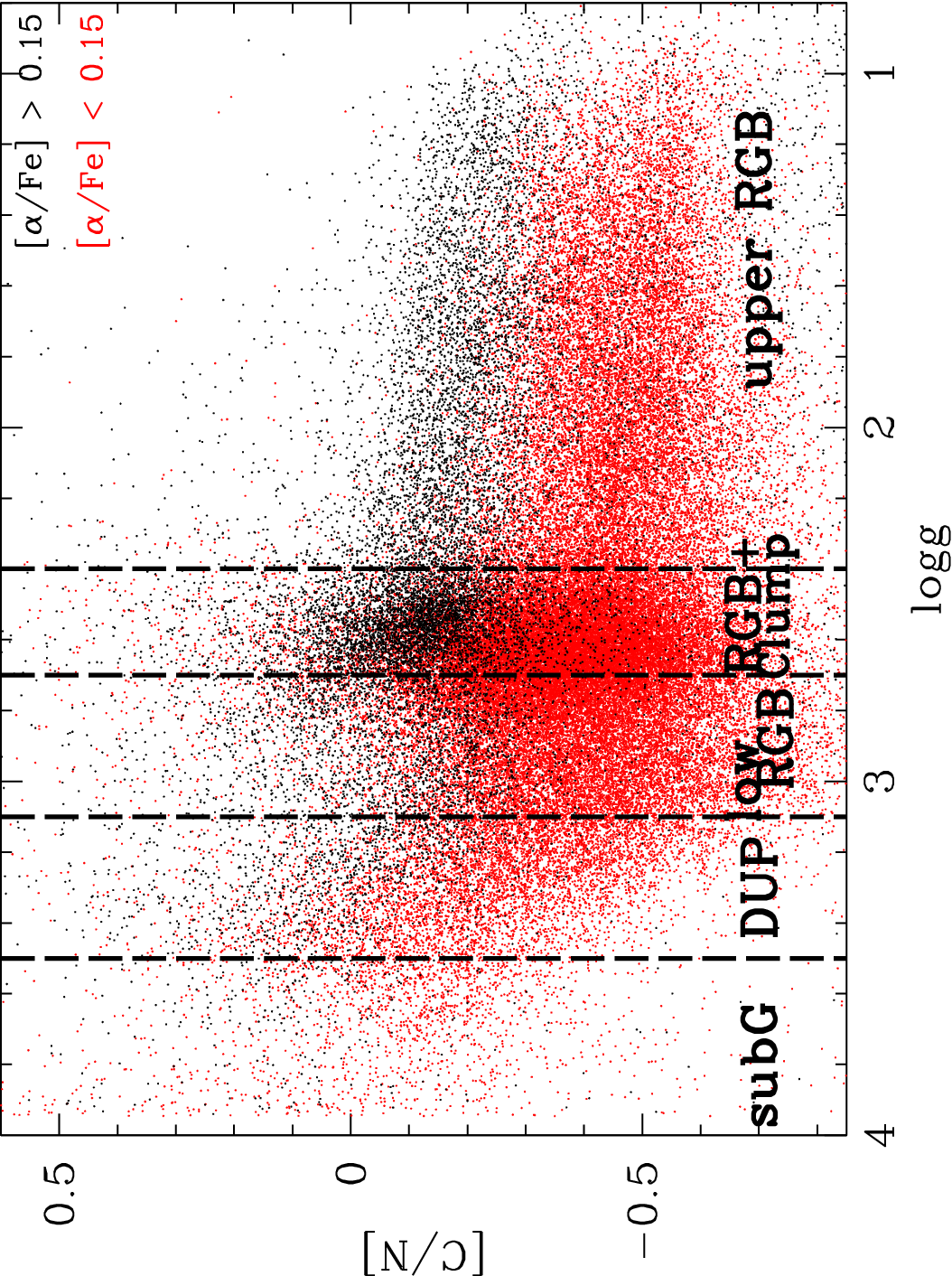}
\includegraphics[width=6cm,angle=-90]{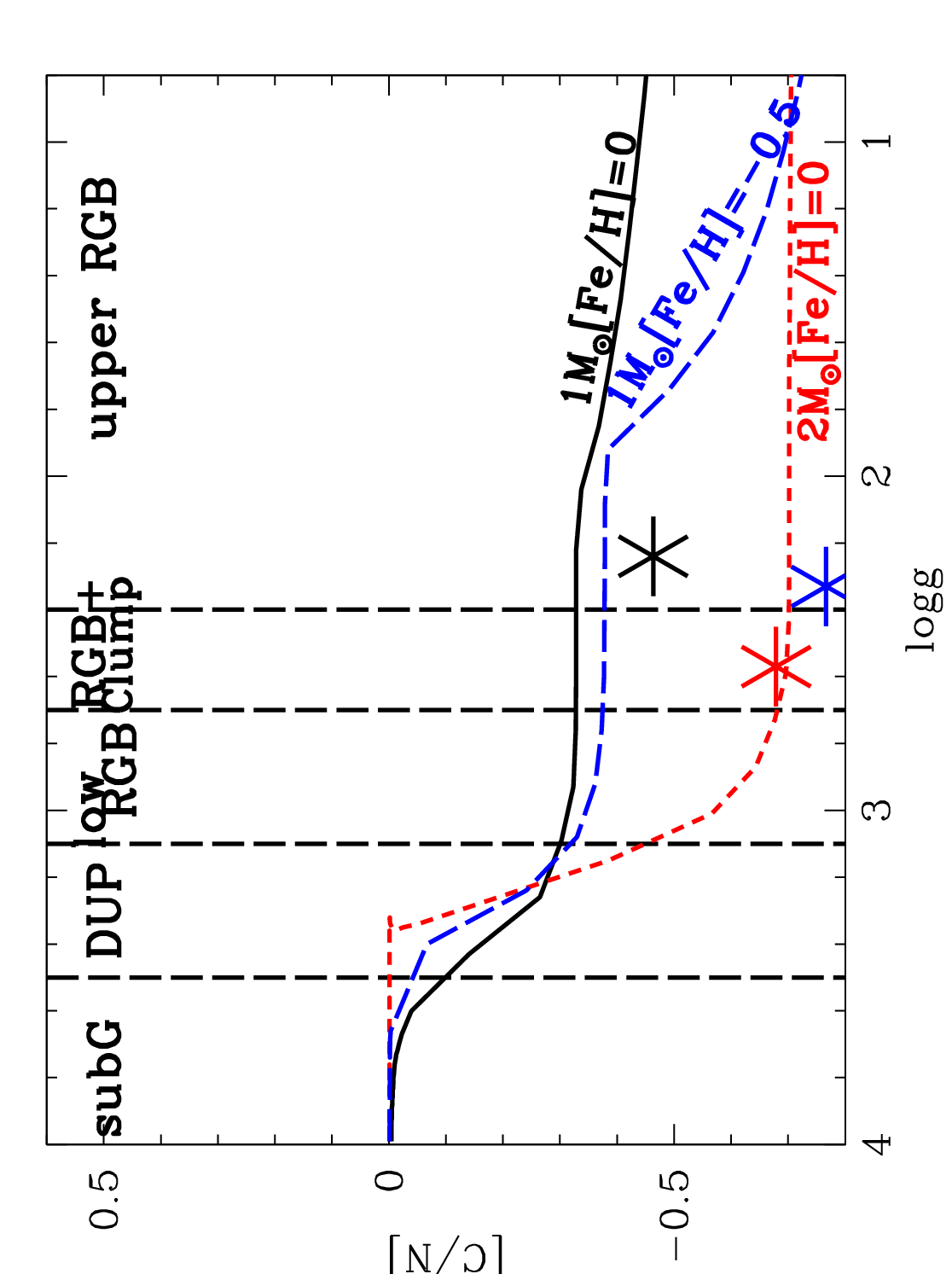}
  \caption{(top panel) [C/N] vs $\log$\,g for the selected sample of  stars from the APOGEE survey with high signal-to-noise spectra, and with $\rm T_{eff} > 4000K$. The stellar sample is colour-coded according to [$\alpha$/Fe], with red points having [$\alpha$/Fe] $< 0.15$ (predominately thin disk) and black points having [$\alpha$/Fe] $> 0.15$ (predominately thick disk). Our empirical stellar evolutionary phases, correlated with stellar surface gravity, are indicated by vertical dashed lines. The stages are sub-giants ("subG"), dredge-up phase ("DUP"), lower red giant branch ("lowRGB"), red giant branch and red clump ("RGB+Clump") and upper red giant branch ("upper RGB").  The lower panel repeats the stellar evolutionary indications, and in addition shows predicted theoretical [C/N] vs $\log$\,g relations for three different masses and metallicities, as indicated, with the models being from \citet{Lagarde2012}. The predicted locations of the red clump are shown by asterisks. The consistency between the data and the model trends is discussed in the text. }\label{fig:CNvslogg}
\end{figure}

The resulting subsample is presented in Fig.~\ref{fig:CNvslogg}, which illustrates the relationship between the stellar [C/N] abundance ratio, the stellar evolutionary state, via log\,g, with a further indication of [$\alpha$/Fe] ratio, divided into two regimes, high ([$\alpha$/Fe]$> 0.15$) and low ([$\alpha$/Fe]$<0.15$). This division is consistent with the definitions observed in other APOGEE studies \citep{Nidever2014,Anders2014} and corresponds to the Galactic thick disk and the Galactic thin disk respectively (cf  Fig.~\ref{fig:alphafe}). The distribution function observed is manifestly inconsistent with a mono-abundance population in the [$\alpha$/Fe] - [Fe/H] plane, as already observed by other independent studies \citep[e.g. ][]{Fuhrmann2011,Adibekyan2011,Lee2011,Bensby2014}. In the figure we can observe the different evolutionary status of the stars and the relationship with the C/N ratio. Since the pioneering work of \citet{Iben1965}, standard stellar evolution theory predicts that after leaving the main sequence stars evolve up the sub-giant (subG) branch. It is expected up to this point that the surface abundances are unchanged since formation for normal single stars. Things change, however, as the star starts its evolution up the red giant branch (RGB).  Model expectation is that the first dredge-up occurs, bringing some of the material  synthesized during  main sequence evolution up to the surface. In particular, this dredge-up enhances the nitrogen surface abundance at the expense of carbon. This expectation is supported by Fig.~\ref{fig:CNvslogg}, where we observe an obvious systematic  decrease of the [C/N] ratio during and after the first dredge-up phase. Moreover, it is well established that the amount of newly-synthesised material mixed to the surface is a function of stellar main-sequence mass. Hence, it is expected that the more massive is a star, the deeper is the dredge up, with more of the nuclear products brought to the surface. In other words, the more massive is a star the lower will be its post first dredge-up C/N ratio. Note also that there is a second-order dependence of the depth of the first dredge-up on stellar metallicity (see \citet{Charbonnel1994} for a fuller discussion). After first dredge-up, as the star evolves along the RGB, more mixing occurs, decreasing further the C/N ratio \citep{Gratton2000, Martell2008}. This observational evidence for non-canonical RGB mixing, that is a process that is not driven by convection, is consistent with the thermohaline mixing mechanism for changes in elemental composition with stellar luminosity \citep{Charbonnel2007}.

To take into account the several effects of stellar evolution on C/N ratios we divide the sample into its five main evolutionary stages (see Fig.~\ref{fig:CNvslogg}): the sub-giant branch ("subG") for which we assume no or little change of the atmospheric elemental composition since the time of star formation; the first dredge-up phase ("DUP"); the lower red giant branch ("low RGB") - which shows the atmospheric elemental abundance composition as modified by the first dredge-up;  the red clump luminosity region of the RGB, including the red clump itself ("RGB + Clump"); and the upper RGB ("upper RGB") - where more CN processing from non-canonical processes may occur.

Note also that for the following discussion we avoid the first dredge-up range because it contains a mix of stars at different stages of their first dredge-up. We also avoid the region defined by $2.4 < \log g  < 3.1$ because there is a contamination by stars belonging to the red clump. Visual inspection of the figure in the range of the clump ("RGB + Clump") suggests that the C/N ratio range is larger than for the upper RGB. This would be surprising, since the C/N ratio is not expected to change after the RGB tip has been reached. This appearance is not real, but is an artefact due to 2 factors: i) the number of sample stars is much larger in the clump than in the upper RGB, which visually enhances the scatter on the plot; ii) the region contains an uneven mix of RGB stars, with relatively high C/N ratios, and clump stars, with lower C/N ratios.

The discussion above demonstrates that [C/N] ratios in the red giants of the Galaxy  observed by APOGEE are predominantly determined by the star's evolutionary state,  and that we can use this ratio to infer the distribution of stellar properties, notably their relative masses. Although our discussion primarily is based on a star to star comparison, it relies on two main requirements: i) The observed C/N ratios are internally on a consistent scale and do not suffer from any dominant bias correlated with stellar parameters ii) the C/N ratio of most stars at the time of their formation  was close to Solar, [C/N]=0. We now demonstrate how and when those requirements can be justified and validated.
 
\subsection{Measurement random errors}\label{sec:biases}

According to \citet{Holtzman2015}, random uncertainties on C and N abundances are respectively 0.06 and 0.07dex. Systematic uncertainties will also be present - see below - but do not restrict the differential analysis we implement for the following reasons:
\begin{itemize}
\item We restrict our analysis into specific evolutionary stages.  This minimises a bias with stellar parameters by restricting to samples that have similar T$\rm{eff}$ and $\rm \log g$. Regarding metallicity, APOGEE data cover a large range. While our discussions remains insensitive to bias when comparing data at a given metallicity, we rely on the careful checks and calibrations from the APOGEE team and assume that the broad trends as a function of metallicity are real. 

\item We select only good quality data (i.e. data with SNR $>$ 100) to ensure that the measurement of line intensities is homogenous for all stars with similar stellar parameters.  

\item {\bf  We compute the star to star dispersion for [C/Fe] at a given metallicity and we obtained 0.08, 0.07, and 0.05 for subgiants, lower RGB and upper RGB stars of the thin disk, while we obtained 0.10, 0.09, and 0.06 respectively for the thick disk stars . Those values can be compared to the average of individual errors on C abundance of 0.05.  We conclude that the total scatter in [C/Fe] at any [Fe/H] and selected evolutionary stage is comparable to the quoted observational errors. As illustrated in Fig.~\ref{fig:CNFevsFe_subG}, similar results are seen for [N/Fe] }

\item It is recognized that errors in stellar atmosphere modelling (such as using 1D vs 3D/hydrodynamical models) can be more pronounced at lower metallicity (e.g.\citet{Magic2013}). C and N are measured in APOGEE via molecular features (respectively CO and CN). Therefore, using the fact that C and N abundances rely on lines forming in the same region of the atmosphere (the coolest part), taking their ratio should then naturally minimise any metallicity (and other stellar parameters) effects due to error in spectral modelling. 
\end{itemize}
{\bf 
To address in particular this last statement, we have completed some tests using the molecular features which are used for the measurement of C, N and O in the APOGEE data (namely OH, CO, and NH). For these, we compare the equivalent widths of each single line of those molecules for a given typical parameter of the APOGEE sample ($\rm T_{eff}=4500K, \log g=2.5$, and solar abundances)  and compare them with those obtained applying the main sources of astrophysical parameter random errors relevant to C and N measurements, i. e. $\rm T_{eff} +250K, \log g+0.5, and [O/Fe]+0.1$, with this latter source of uncertainty being especially relevant for molecular equilibrium concerns. In the upper panel of Fig.~\ref{fig:deltaabu}, we can observe that changes in stellar parameters change the equivalent widths of molecular lines as expected. While we refer the reader to the work of \citet{Tsuji1973} for a deeper understanding of the reasons for those changes, the net result is that both CN and CO molecules show very similar changes in their lines strength as a function of temperature and surface gravity changes. However, changes in surface gravity affect differently the OH and  the CN molecules. Moreover, as illustrated in the third panel of the same figure, O abundance changes also affect OH and CN. In contrast, it very little affects the CO molecule strength, because we consider here $\rm C/O < 1$, so that the number of CO molecules is bounded by the number of free C atoms, and not of O atoms. To consistently evaluate the propagation of all those errors, we thus compute the changes in abundances in the bottom panel of Fig.~\ref{fig:deltaabu}, considering that O, C, and N abundance measurements are derived respectively from OH, CO and CN. It is remarkable to note in that figure that all the changes in parameters result in changes in abundances in the same direction and similar amplitude. Therefore, we conclude that systematic measurement errors on C and N abundances are expected to partially cancel out when considering their ratio.\\

 One may wonder what would be the net effect for other combinations of parameters changes. Actually, we have demonstrated that the O abundance is directly or indirectly the main driver for differential changes between C- and N- bearing molecules, and thus to change in C/N ratios. Therefore, another set of parameters will lead to similar results.\\
 For this test, we have chosen the worst case scenario, meaning that the  errors used in are significantly larger than those published in \citet{Holtzman2015}, and further that those random errors have been added up, while in a more realistic treatment random errors should be randomly distributed, thus reducing further the amplitude of any variations in abundances simulated here.\\
 Including random errors related to noise in a more realistic test will only affect the line to line dispersion, and thus is not expected to change the average abundances.

\begin{figure}
\includegraphics[width=6cm,angle=-90]{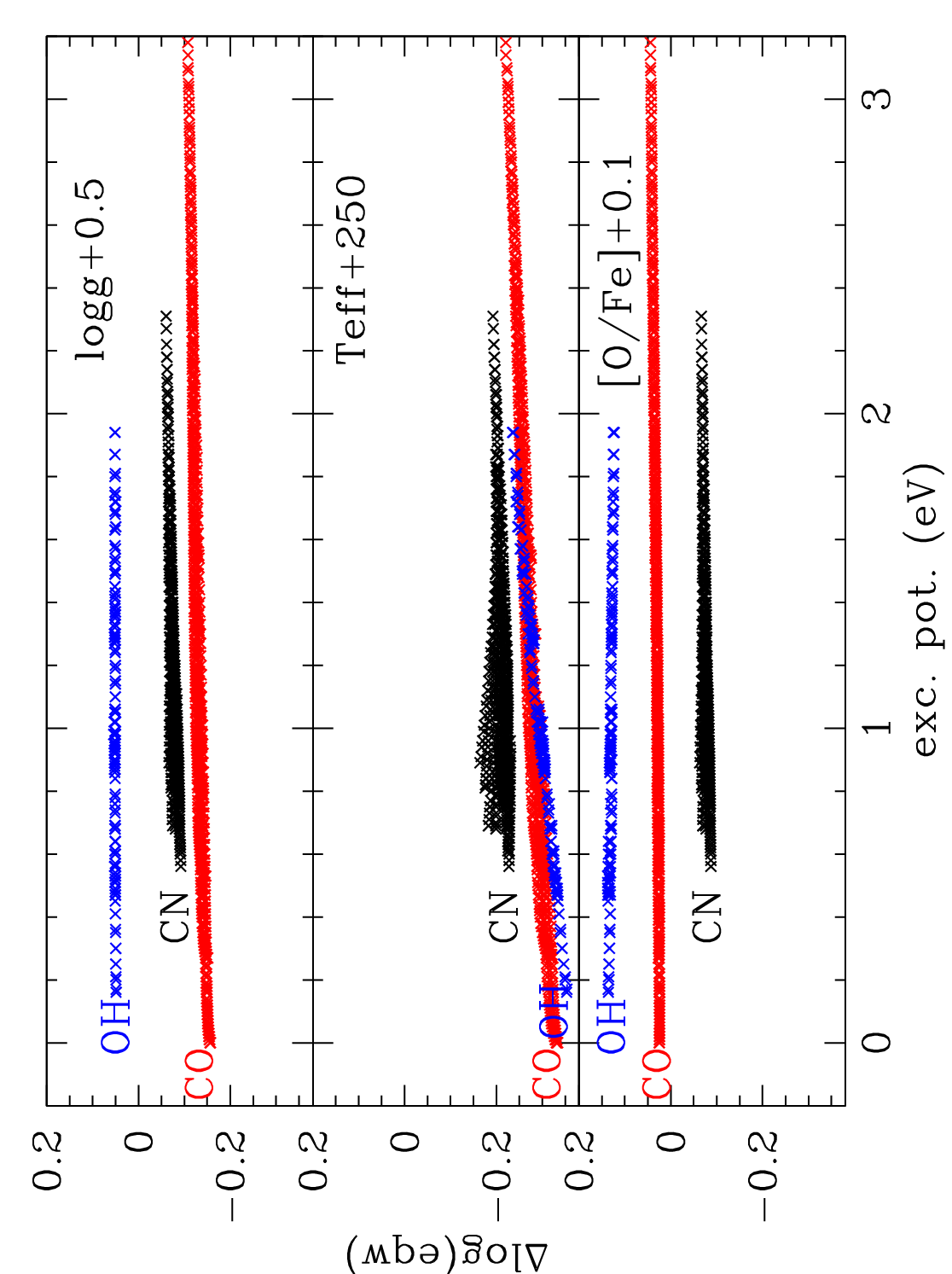}
\includegraphics[width=6cm,angle=-90]{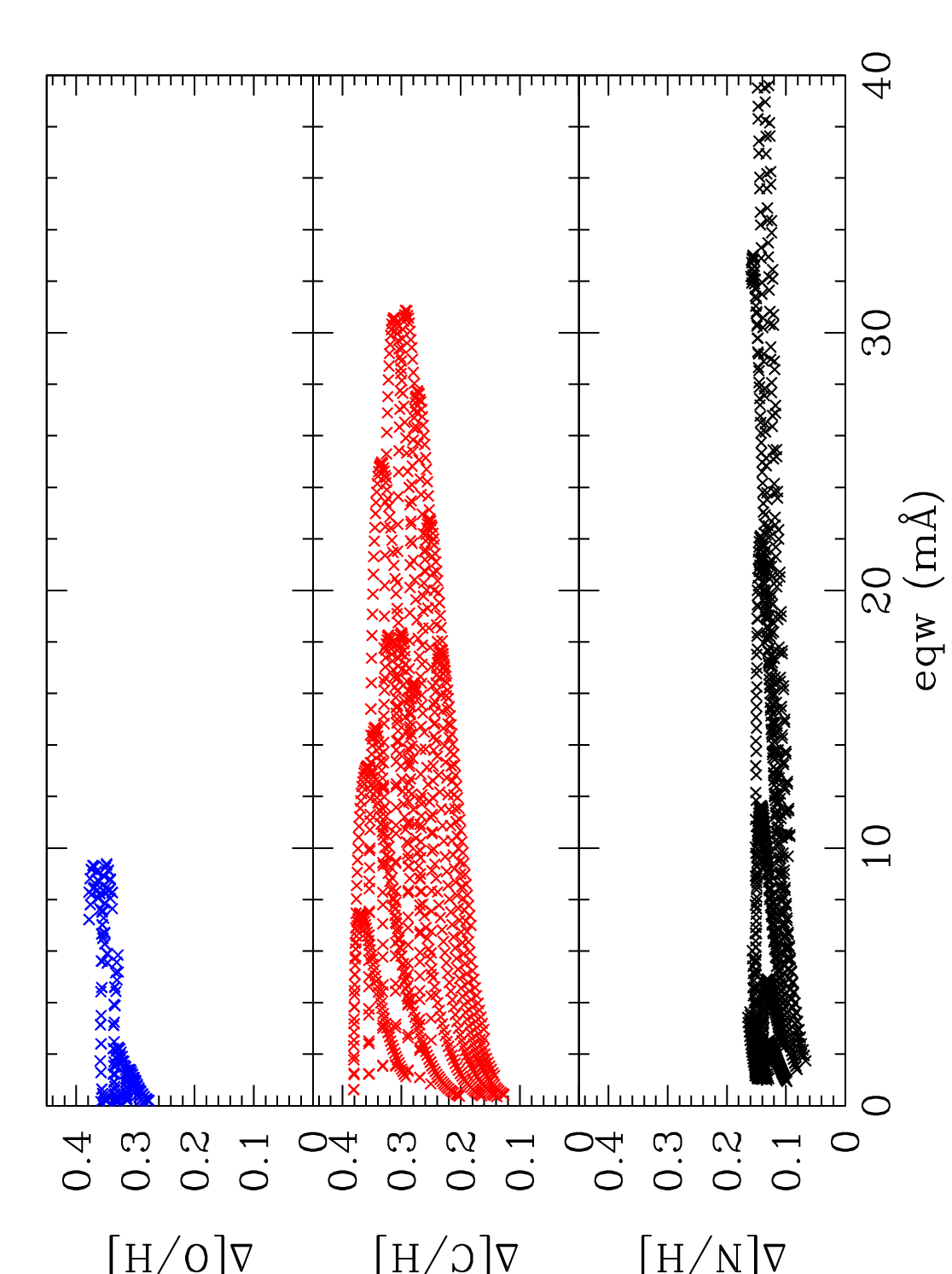}
\caption{Upper panels: Differences in line equivalent widths of molecules present in the APOGEE spectra obtained when changing stellar parameters from $\rm T_{eff}=4500K, \log g=2.0, [O/Fe]=0.0$ to the indicated values. Lower panels: Differences in C, N and O abundances measured when changing simultaneously all the stellar parameters. The overall change in parameters affects the abundances with the same sign and with comparable amplitude, supporting the idea that using those elements as ratios is robust. }\label{fig:deltaabu}
\end{figure}
}
Based on all those arguments, the precision of the C/N ratio is at least as good as 0.1dex at fixed [Fe/H] and evolutionary state, which is sufficient precision to derive quantitative properties of the Galactic disks, which we do in the next section. 

\subsection{Measurement systematic errors}\label{sec:pristine}

\begin{figure}
\includegraphics[width=6cm,angle=-90]{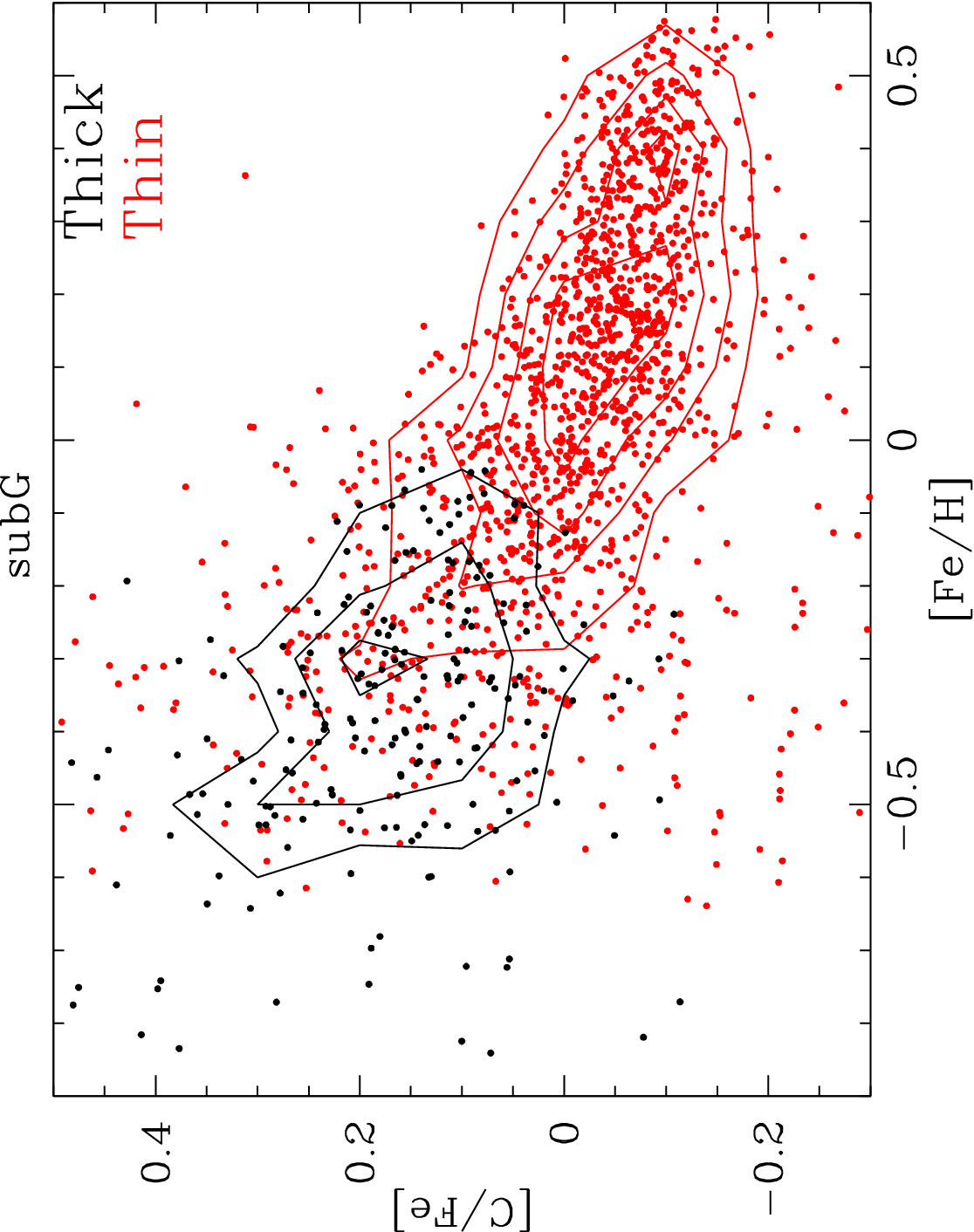}
\includegraphics[width=6cm,angle=-90]{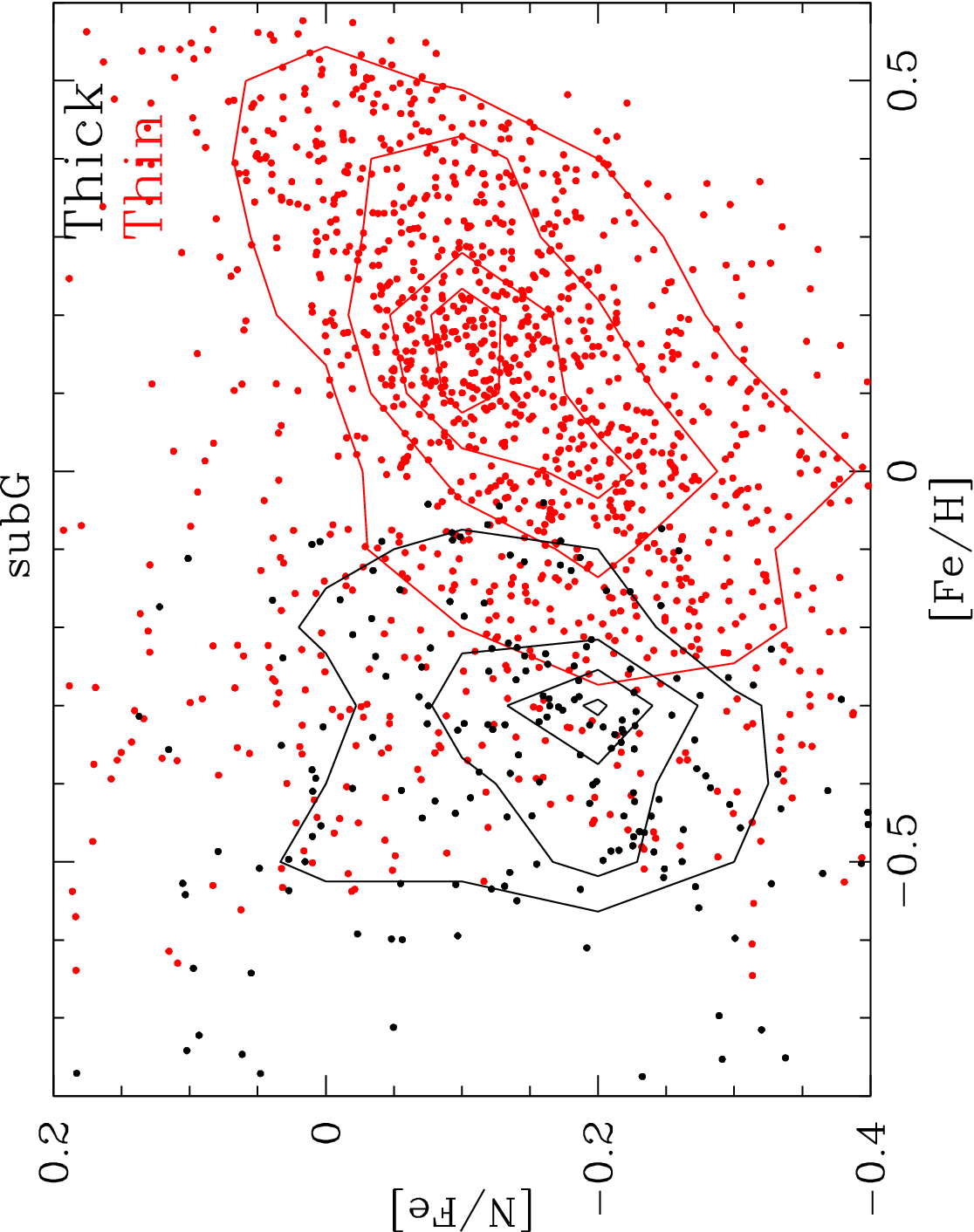}
\includegraphics[width=6cm,angle=-90]{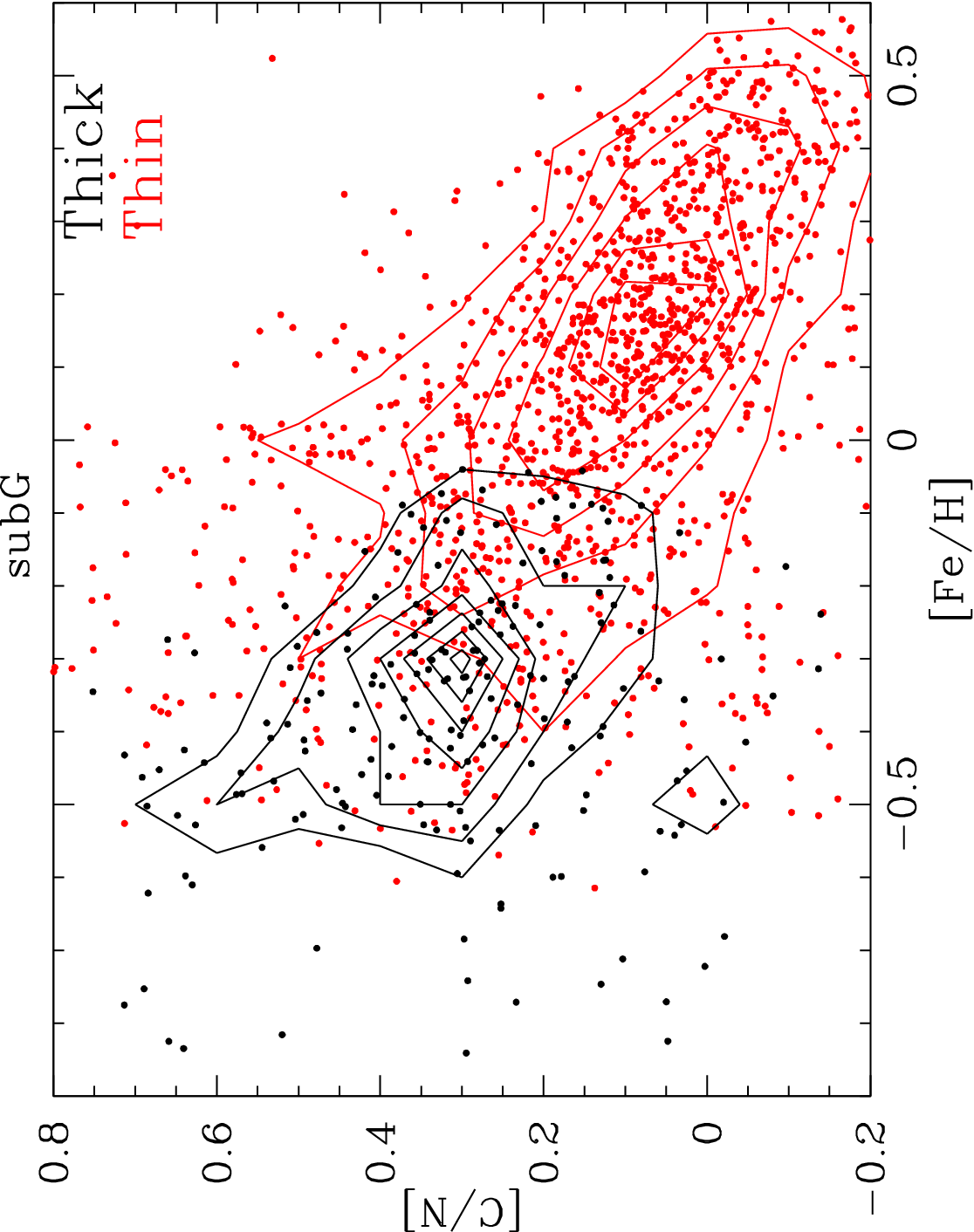}
  \caption{Testing data scatter and zero points with pre-dredge-up subgiants. The three panels show [C/Fe], [N/Fe] and [C/N] as a function of metallicity for the APOGEE subgiant stars. There is a continuous decrease of C/N ratio with metallicity, consistent with plausible expectations from Galactic chemical evolution. Both low [$\alpha$/Fe] (thin disk) and high [$\alpha$/Fe] (thick disk) sub-giant stars have consistent values in the region of [Fe/H] where both populations are present. The definition of thick and thin disks is described in the text.  We also note that [N/Fe], and consequently [C/N] is not equal to 0 at solar metallicity, suggesting a zero-point offset in N abundances.}\label{fig:CNFevsFe_subG}
\end{figure}

The discussion above, which supports the published estimates for precision, does not hold for zero-point accuracy. 

Comparison of the [$\alpha$/Fe] vs [Fe/H] distributions in APOGEE (Fig.~\ref{fig:alphafe}) with those of Gaia-ESO \citep[esp their Fig 3]{Mikolaitis2014} suggests small systematic scale differences at the 0.1dex level in element ratio and/or Fe-abundance zero point. These do not affect the discussion here.

Checking the calibration of C and N abundances is harder, since, as the discussion above explains, these change during a star's evolution. However, there is available a relatively small sample of sub-giants. These stars have not yet experienced dredge-up, so their surface abundances reflect the abundance of the inter-stellar medium from which they formed.  Fig.~\ref{fig:CNFevsFe_subG} shows the ratios of [C/Fe], [N/Fe] and [C/N] vs [Fe/H] for the sub-giant sample. One's expectation for the appearance of these figures is a function of one's astrophysical preconceptions and preferred chemical evolution model. The [C/Fe] vs [Fe/H] relation decreases slowly and smoothly with increasing metallicity, passing close to Solar at Solar [Fe/H]. This is astrophysically plausible and already observed by \citet{Nissen2014}, suggesting the carbon abundance calibration is not unreasonable. The [N/Fe] vs [Fe/H] relation is roughly constant below Solar, with a systematic increase above solar metallicity.  This evolution looks plausible as one can expect the nucleosynthetic products of low mass stellar evolution, in particular N production, to become increasingly significant compared to supernovae products, which have low N production, with increasing time. We do not of course know that the super-solar stars are indeed relatively young, so this remains a consistency statement, not a proof. As a striking example, Fig.~\ref{fig:CNFevsFe_subG} clearly shows that the $\rm[N/Fe]$ value at solar metallicity does not match the Solar ($0$) value for the bulk of the stars, as already observed by \citet{Israelian2004}, highlighting the preliminary calibrations for this element in the current APOGEE data. There is a zero-point offset in the [C/N] ratio of approximately 0.2dex, apparently dominated by the offset in [N/Fe]. Although this does not affect the conclusions of this study, which are based on purely differential comparisons, we apply an constant offset correction in the N abundance of +0.2 for all the APOGEE sample for the rest of the paper.

The [C/N] ratio varies smoothly between low metallicity sub-giants and high metallicity sub-giants. This range, if a calibration artefact, could affect this study. However, we note that the [C/N] ratio at given metallicity is uncorrelated with [Fe/H] (lowest panel of Fig. ~\ref{fig:CNvslogg}), at least according to models, which suggests that the trend is not a calibration error, and in any case will not affect our conclusions below, which depend entirely on the different behaviour of [C/N] vs [Fe/H] as a function of [$\alpha$/Fe].  More importantly, the change in [C/N] after the stars have undergone their first dredge up is several times larger than the scatter in the sub-giant range, and is an order of magnitude larger than any systematic dependence of [C/N] on [$\alpha$/Fe]. In summary, for the purposes of this study, we may be confident that the apparent variation of the [C/N] ratio in the sample is not a calibration artefact, but is dominated by internal nuclear processes and stellar evolution.

For convenience, in the following discussions and figures we recalibrate the APOGEE data by applying an offset of +0.2 in N abundances to match the solar abundance.  This offset has no effect on any of the discussion below unless noted, and we do not consider our adopted value of +0.2dex a precise or recommended number.

\section{The thick and thin disks are discrete in [$\alpha$/Fe] vs [Fe/H]}\label{sec:thickthin}

\begin{figure}
\includegraphics[width=6cm,angle=-90]{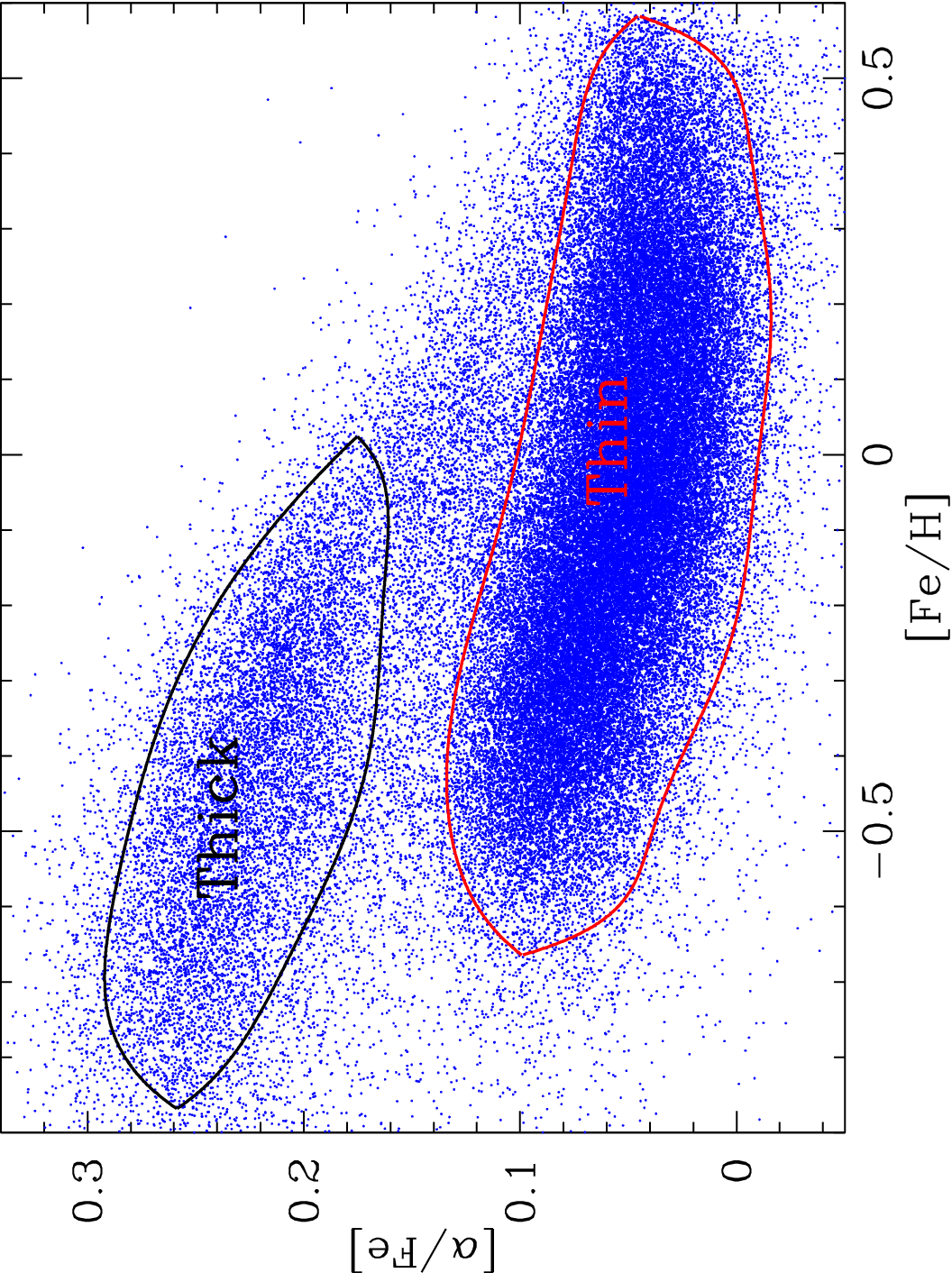}
\includegraphics[width=6cm,angle=-90]{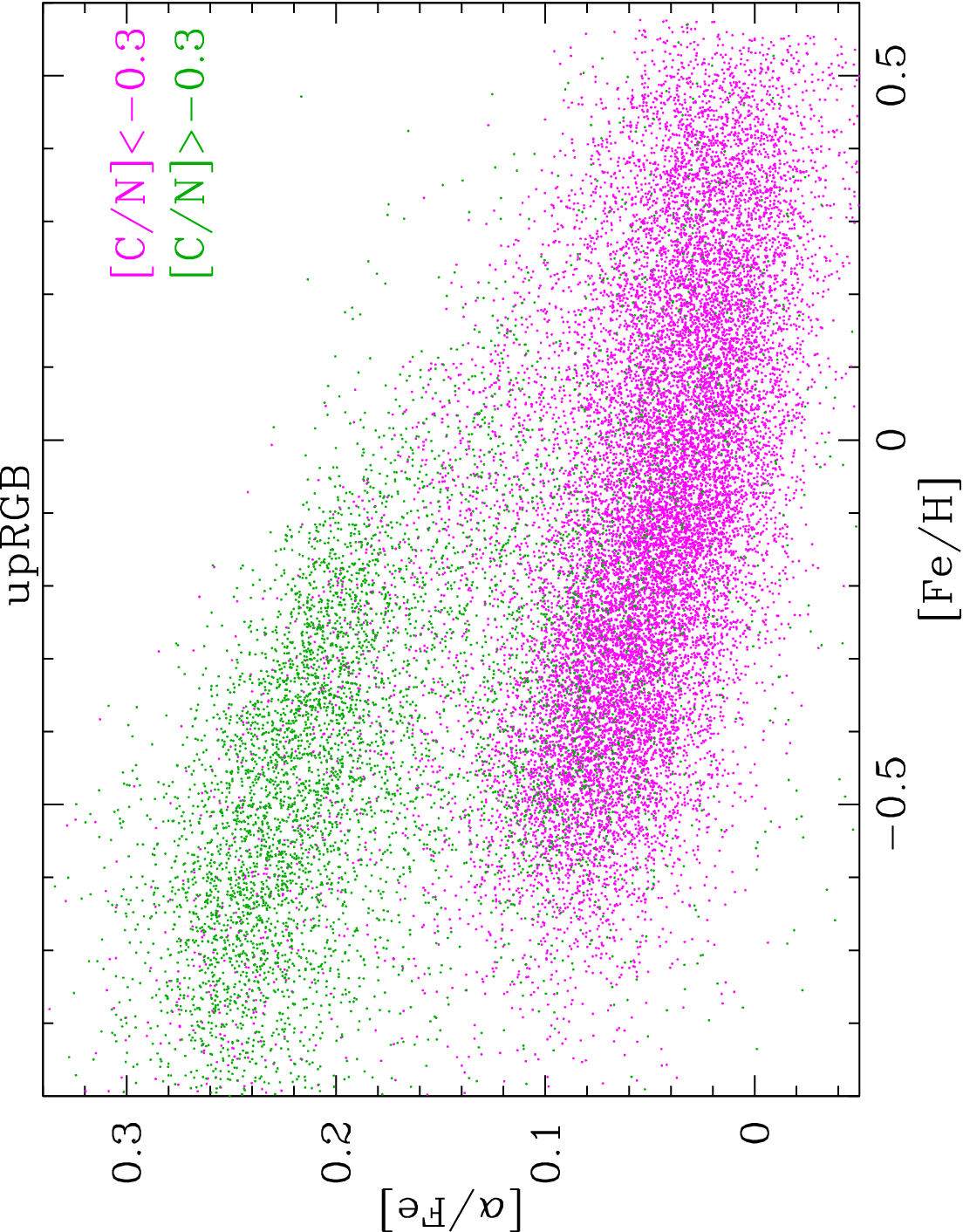}
  \caption{(top panel) The distribution of $\rm[\alpha/Fe]$ as a function of metallicity for APOGEE upper RGB  stars. The distribution into two broad maxima, with a clear minimum in number density between them is very evident. We identify the upper distribution within the density contours as thick disk, and the lower distribution within the contours as thin disk.  In the bottom panel data for the upper RGB stars are color-coded in two bins:  [C/N]$< -0.3$, and [C/N]$> -0.3$. This discrimination corresponds to a combination of intrinsic stellar properties, especially main-sequence mass, and stellar evolution, dredge-up and other mixing events, and not to Galactic spatial distribution or kinematics. The [C/N] selection clearly correlates tightly with [$\alpha$/Fe] confirming that the two identified stellar populations do form discrete populations.}\label{fig:alphafe}
\end{figure}

It has been established by many studies that the $\alpha$-elements provide a discrimination between the thick and the thin disk, with the thick disk having systematically higher $[\alpha$/Fe] abundance rations than does the thin disk at given [Fe/H] (\citet{Mikolaitis2014}). Although questioned by \citet{Bovy2012} in their analysis of low-resolution SDSS-SEGUE data, the present high-resolution SDSS-APOGEE data show a very clear distinction, with strikingly discrete distributions. This is shown in Fig.~\ref{fig:alphafe}. The distribution is into two broad maxima, the upper of which we identify as the thick disk, the lower as the thin disk. We emphasise that, for present purposes, the contours chosen are indicative. We make no attempt here to fine-tune the selection, with in particular the extension of the high-$\alpha$ group all the way up to [Fe/H]=0 being suggestive rather than robust. Rather the whole point is that the distinction between the dominant two populations is so clear. This bimodal distinction is not an arbitrary density-labelling, as we show in the lower panel, which shows the high correlation between the [$\alpha$/Fe]-based population definition and the [C/N] ratio. Rather the stars identified by the density contours form two distributions which are clearly different in their stellar evolutionary histories, and in their present-day age distributions.   We consider that evidence further in the remainder of this paper.

An additional advantage of using the [C/N] ratio is that  this new population indicator relies on molecular features, which can be observed in relatively low resolution spectra, unlike the case with $\alpha$ elements. Thus C and N provide an excellent and robust spectroscopic complement to population allocation of thin disk and thick disk red giant stars. 

We do not here attempt to analyse population numbers, or selection effects. Clearly at the metal-poor end the APOGEE selection bias is significant. In the region where both populations are evident, essentially for stars with $-0.6 \leq {\rm[Fe/H]} \leq -0.2$, there is no obvious selection effect which can generate a tight correlation between two group of elements with very different properties.

 The distribution function observed is manifestly inconsistent with a mono-abundance population in the [$\alpha$/Fe] - [Fe/H] plane. We do note also that there does seem to be a real excess over random scatter of stars near solar metallicity with high [$\alpha$/Fe] values. We discuss these more below.

\section[]{Carbon and Nitrogen along the RGB: determining ages for the thick and thin disks}

The $\alpha$-elements retain a memory of the material from which a star has formed.  Carbon and nitrogen, on the other hand, convey age-dependant information on the evolutionary history of the individual star. We now exploit those distinct properties of stars to discuss simultaneously the history of the thin and thick disks, and the imprints of stellar evolution.
 
\subsection{The lower RGB stars: Star formation histories and relative ages of the thin and thick disk}
\subsubsection{Observations}
We consider first those stars classified in Fig.~\ref{fig:CNvslogg} as in the lower RGB phase of stellar evolution, after first dredge-up. 

\begin{figure}
\includegraphics[width=6cm,angle=-90]{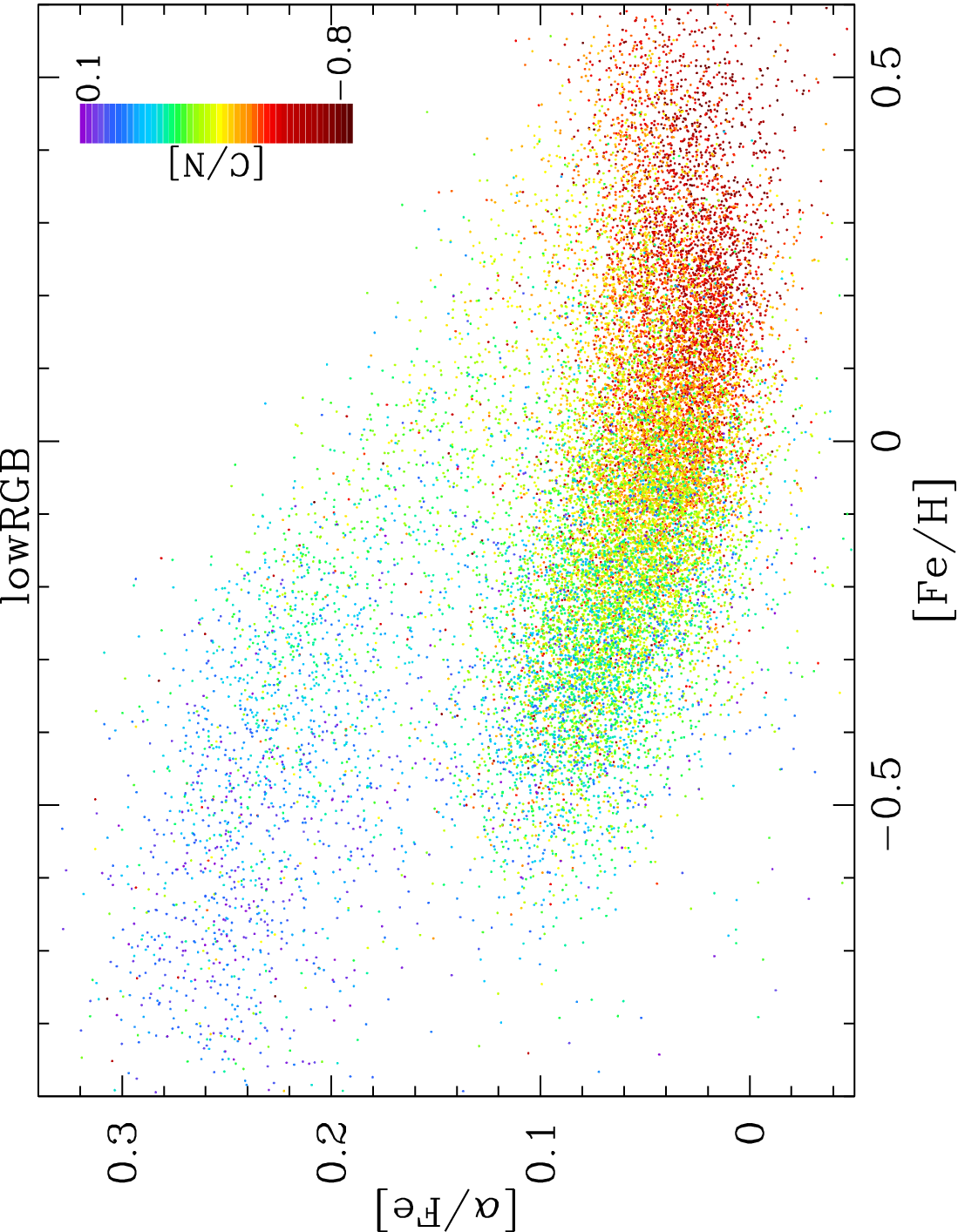}
  \caption{$\rm[\alpha/Fe]$ as a function of metallicity for lower RGB stars. The C/N ratios are color-coded as indicated on the scale bar. We can observe a C/N gradient within the thin disk stars. This is consistent with the  dependence of the first dredge-up on mass and metallicity, implying, unsurprisingly, that more metal-rich thin disk stars are systematically younger than more metal-poor stars. See the text for a more detailed discussion.}\label{fig:alphavsFe_CN_rainbow_lowRGB}
\end{figure}

In Fig.~\ref{fig:alphavsFe_CN_rainbow_lowRGB}, which presents the [$\alpha$/Fe] vs [Fe/H] relation for the lower RGB stellar sample, we have  colour-coded the [C/N] values. The gradient in colours can then be translated into a gradient in [C/N] which is clearly correlated with metallicity in the thin disk stars. As noted above, and as illustrated in the lower panel of Fig.~\ref{fig:CNvslogg}, the value of an observed  [C/N] ratio can be due only to dependencies on two properties of the specific star: a metallicity dependence and a mass dependence. Moreover, a more quantitative examination of figure 5 of \citet{Charbonnel1994} tells us that for stellar masses below $\rm \approx 3 M_{odot}$  it is  stellar mass which dominates the changing [C/N] ratio, while for higher stellar masses it is the metallicity.       

\begin{figure*}
\includegraphics[width=6cm,angle=-90]{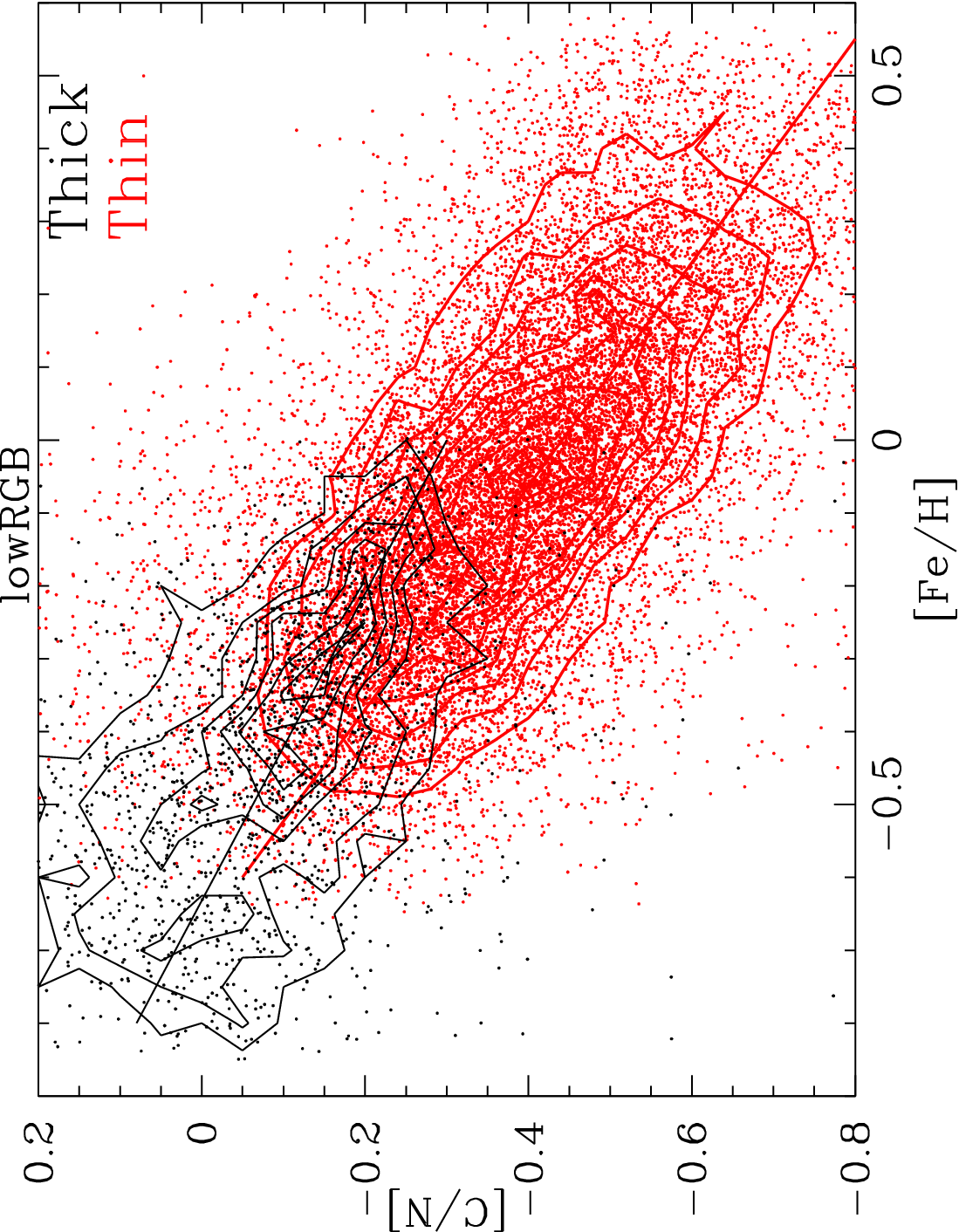}
 \caption{C/N ratios as a function of metallicity for lower RGB stars (dots), colour coded based on the [$\alpha$/Fe] population discrimination as thin disk stars (red) and thick disk stars (black). Density contours are also displayed to guide the eye. We recall that N abundances are zero-point corrected by +0.2 from the original data. The distributions of [C/N] ratios are different between the two populations, showing that they have different age and star formation histories. }\label{fig:CNvsFe_lowRGB}
\end{figure*}

 Fig.~\ref{fig:CNvsFe_lowRGB} presents the distribution of [C/N] as a function of metallicity for lower RGB stars, with the [$\alpha$/Fe] definition selection colour-coded. When comparing thin and thick disks stars, the degeneracy between the mass and metallicity effects on the C/N ratio can be broken where the two disks have an overlapping metallicity range. Therefore, when examining Fig.~\ref{fig:CNvsFe_lowRGB}, one can understand that the thick disk contains in general lower mass stars than the thin disks. Considering the fact that at a given metallicity there is a one-to-one relation between stellar mass and age, we can conclude that C/N ratios represents a clean proxy for time and that the thick disk population formed earlier than the thin disk population.

\subsubsection{Simulation of the [C/N] ratio in lower RGB stars}
\begin{figure*}
\includegraphics[width=6cm,angle=-90]{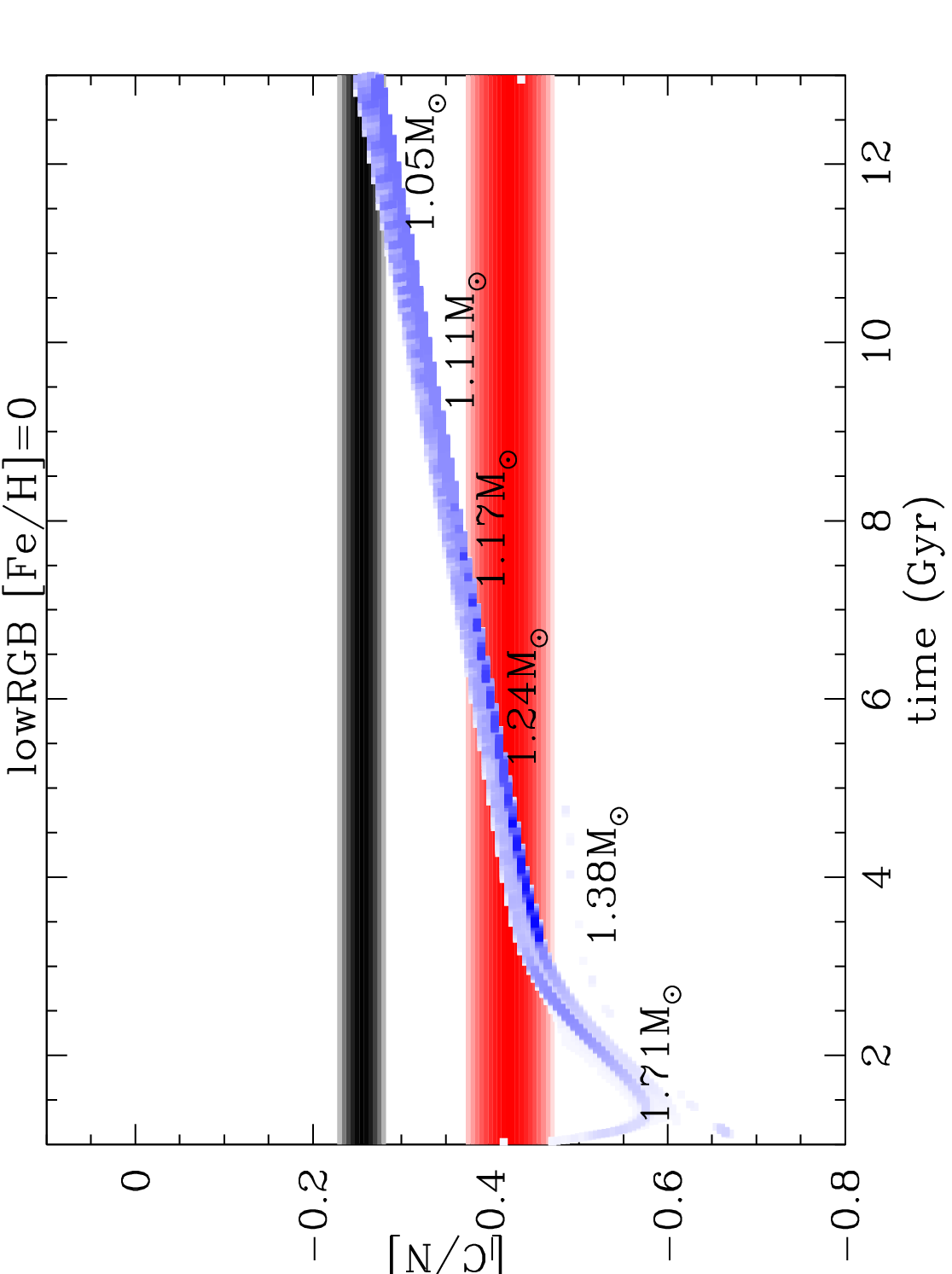}
\includegraphics[width=6cm,angle=-90]{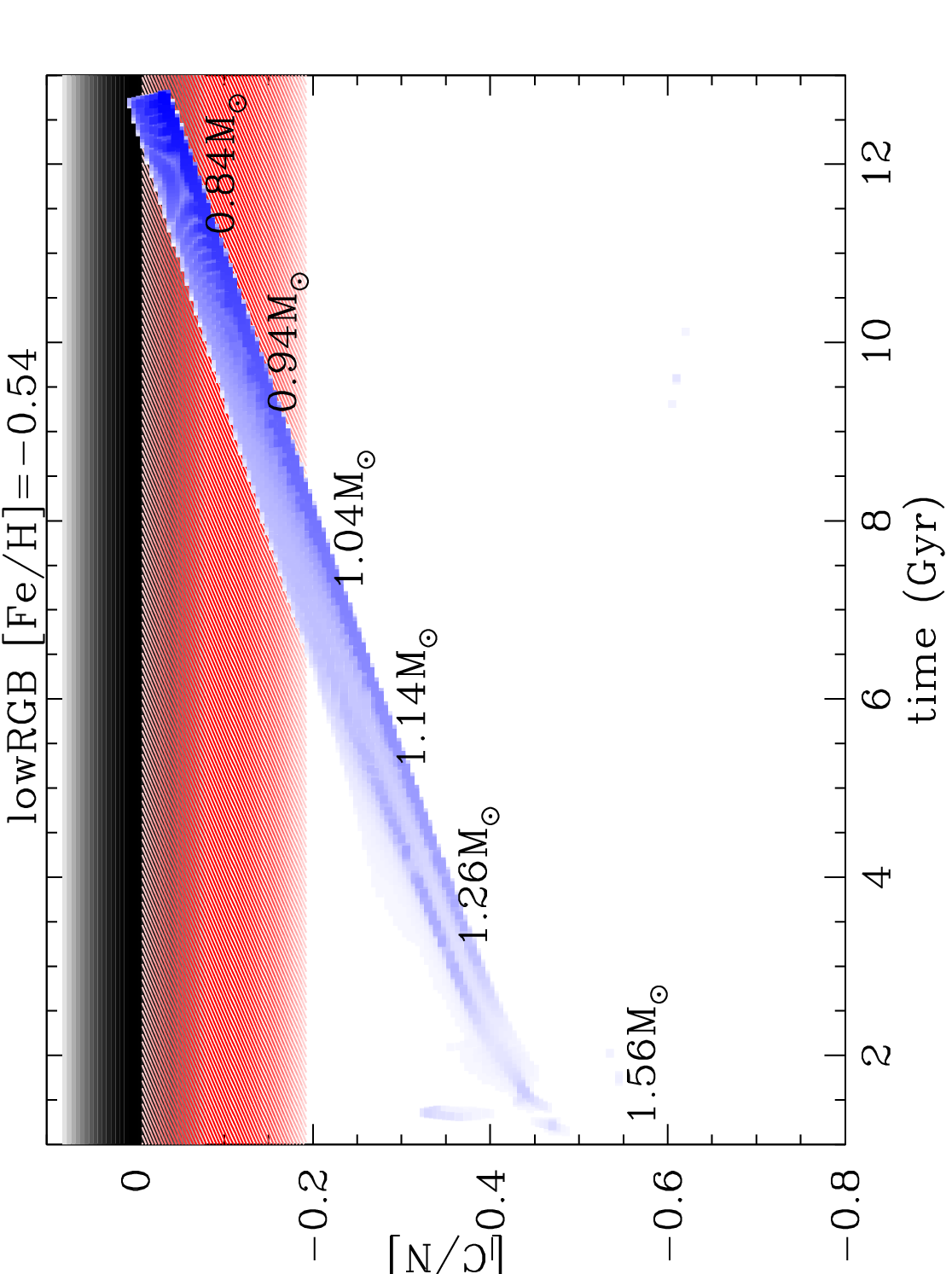}
  \caption{Simulation of the evolution of [C/N] in an RGB star population with metallicity [Fe/H]=0 (left panel) and [Fe/H]=-0.54 (right panel) as a function of time. The masses listed indicate the main sequence mass of the stars dominating in density at a given time and metallicity. The models adopt the stellar grid from \citet{Lagarde2012}. The red and the grey areas represent the values as observed in Fig.~\ref{fig:CNvsFe_lowRGB}  in respectively the thin and thick disk.}\label{fig:CNvstime_lowRGB}
\end{figure*}

To verify and quantify to some extent the evolution of [C/N] ratios with time, we employ here stellar evolutionary models.
It is not trivial to reliably disentangle the effect of mass and metallicity on the C/N ratio to determine ages. Hence, to evaluate more quantitatively the relative ages of the thin and thick disk, we used the stellar model grid of \citet{Lagarde2012} to simulate the evolution of [C/N] ratio with time (Fig.~\ref{fig:CNvstime_lowRGB}). We first linearly interpolate the \citet{Lagarde2012} grid in time and mass with steps of respectively $dt = 10^5$ years and $dM = 0.02M_\odot$. Then, we  derive the density of stars $\rho$ as a function of time and C/N ratio using the following relation:
\begin{displaymath}
\rho(t,[C/N])  =   \int_{t}^{t+\delta t}\int_{CN}^{CN+\delta [C/N]}\int_{\log g_{min}}^{\log g_{max}}M^{-\alpha} dM dCN dt 
\end{displaymath}
where $M$ is the initial star mass, $t$ is the time, C and N are the carbon and nitrogen surface abundances at time $t$,  and $\log g$ is the surface gravity as provided by the stellar evolution models.  We also assume a Salpeter IMF for all the Galactic stars, thus $\alpha=2.35$, which is an acceptable approximation over the very small range in masses relevant here. We applied the same selection definitions based on $\log g$ as in Fig.~\ref{fig:CNvslogg}: thus for low RGB stars $\log g_{min} = 2.7$ and  $\log g_{max} = 3.1$. We chose $\delta t$ and $\delta [C/N]$ to be respectively $\rm 10^7$ years and 0.01 dex as a compromise between numerical limitations and simulation resolution. This resolution as well as linear oversampling of the initial grid, may lead to some uncertainties in the actual star densities, but in the following discussion only the relative evolution of [C/N] as a function of time matters.   We also consider the impact of the initial [C/N] value on the post dredge up value. While the [C/N] value observed in subgiant stars at solar metallicity is compatible with the value in the models, the [C/N] value in the subgiants at [Fe/H]=-0.54 is slightly enhanced, by 0.2 dex (see Fig.~\ref{fig:CNFevsFe_subG} but considering the zero point correction in N) due to enhanced carbon at lower metallicities. Hence, we add this value to our final stellar density value $\rho$ when running the low metallicity models. This should be a reasonable approximation to mimic the dilution of the first dredge up products in slightly [C/N]-enhanced material. We consider the consistency between the model abundances and the observed values further below.

As expected, there is definitely a one-to-one relation between the value of [C/N] and age. What is also striking with this simulation is the tightness of the relation, making the C/N ratio potentially a very robust time indicator, provided the models are a good match to data.  The slope of the relation in Fig.~\ref{fig:CNvstime_lowRGB} is such as to allow age determinations with formal precision of order 1-2 Gyr. Hence we confirm that thick disk stars are older than thin disk stars - according to the models.

\subsubsection{Simulation uncertainties}
\begin{figure}
\includegraphics[width=3cm,angle=-90]{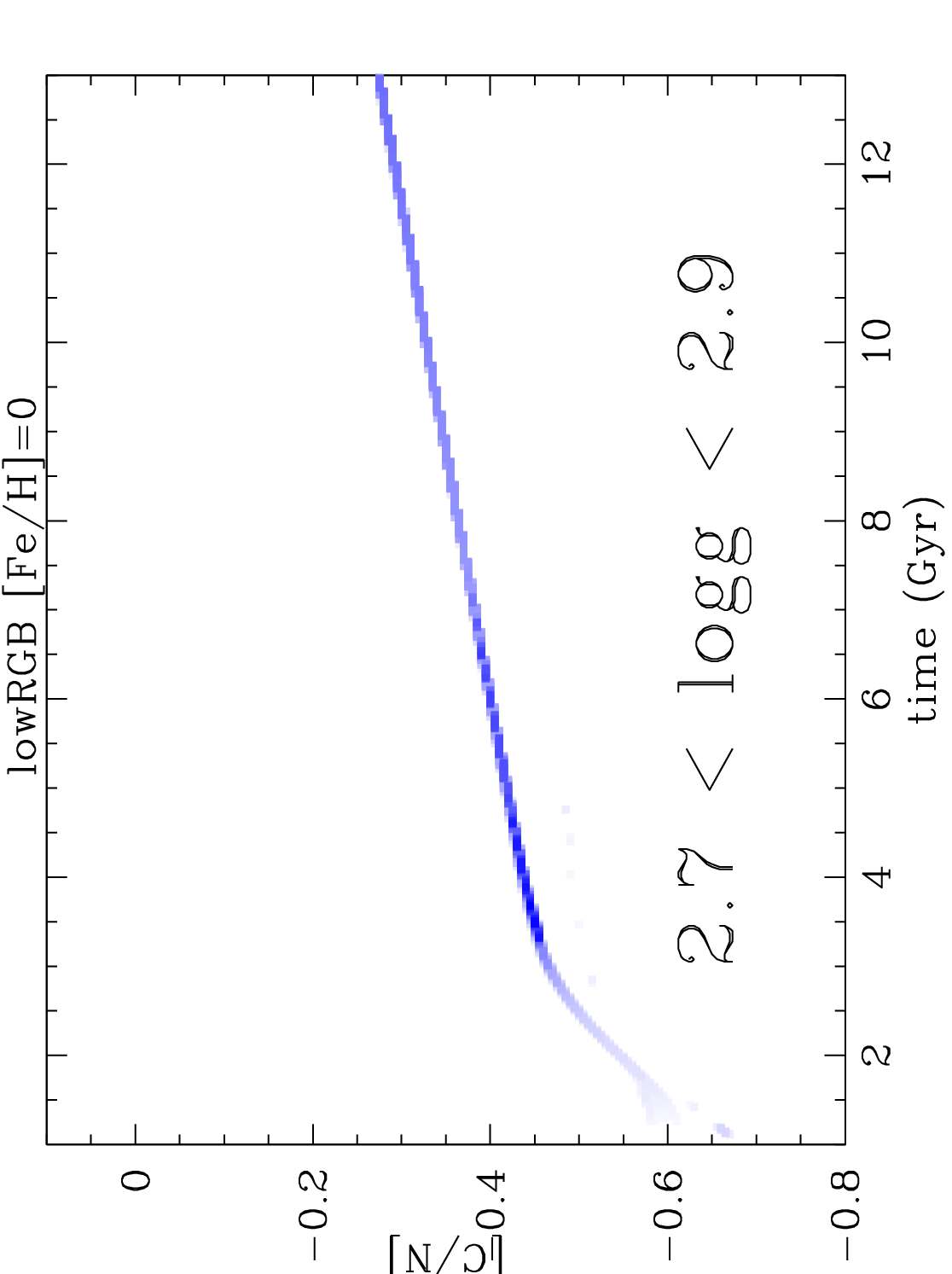}
\includegraphics[width=3cm,angle=-90]{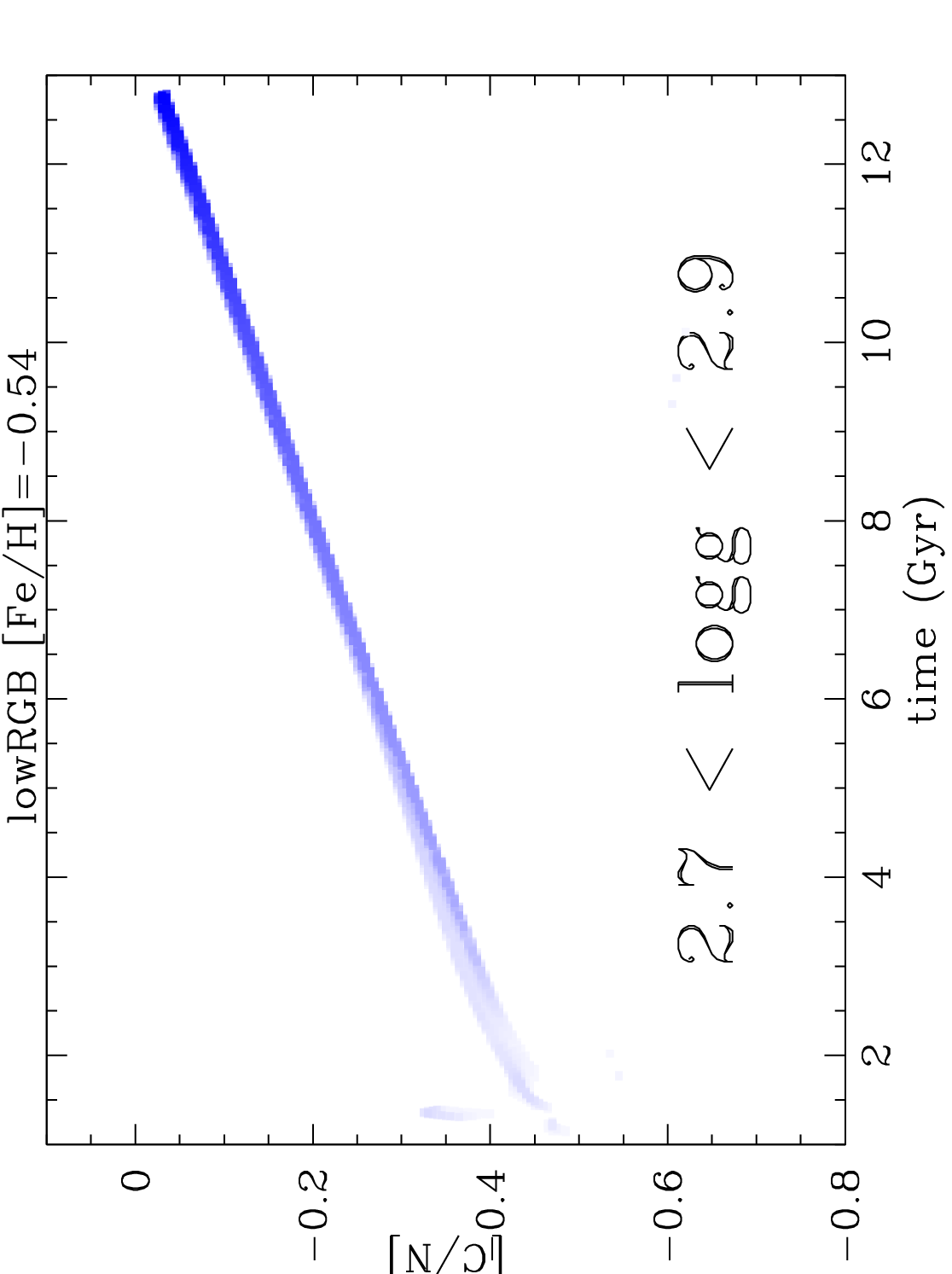}
\includegraphics[width=3cm,angle=-90]{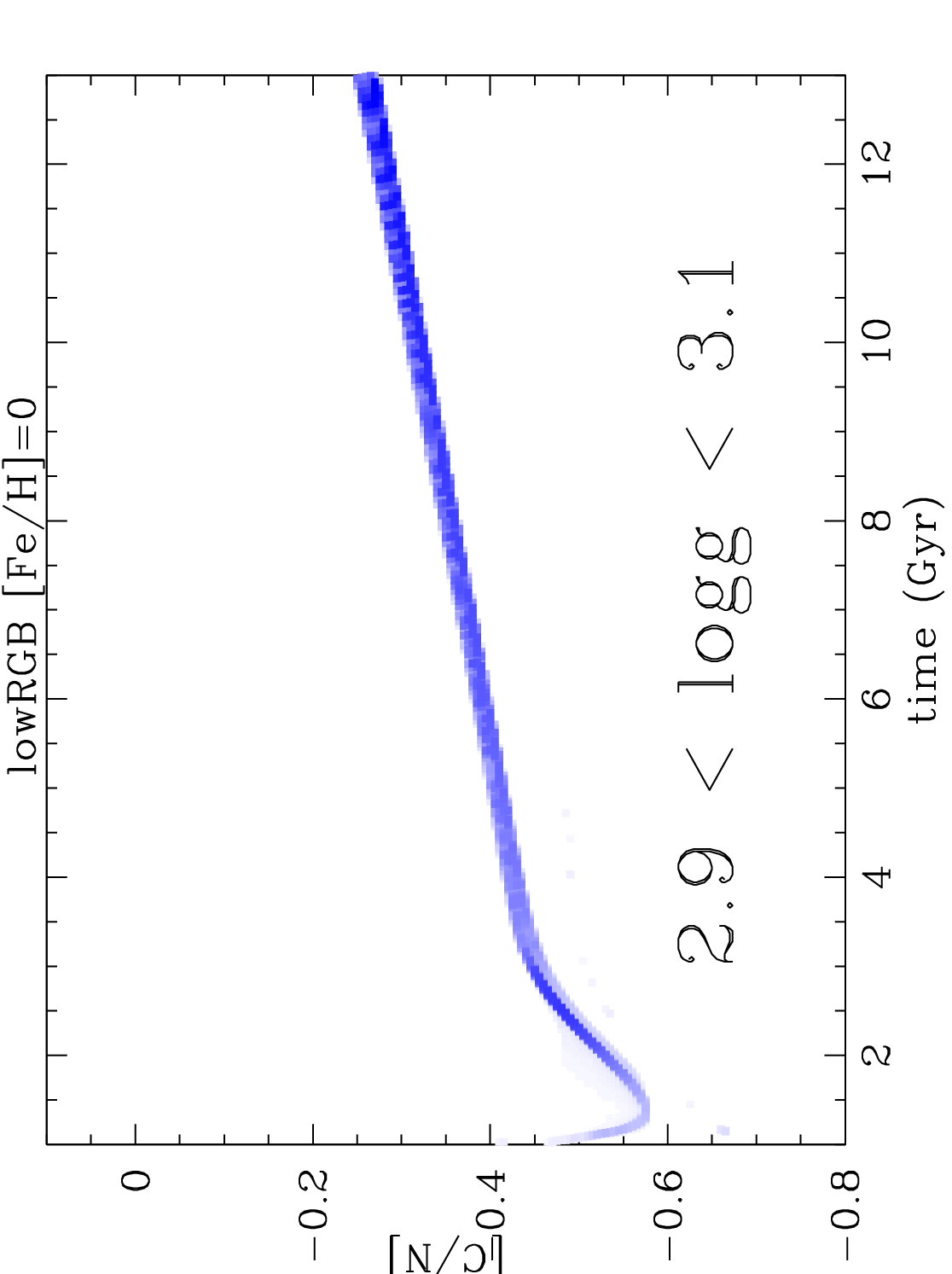}
\includegraphics[width=3cm,angle=-90]{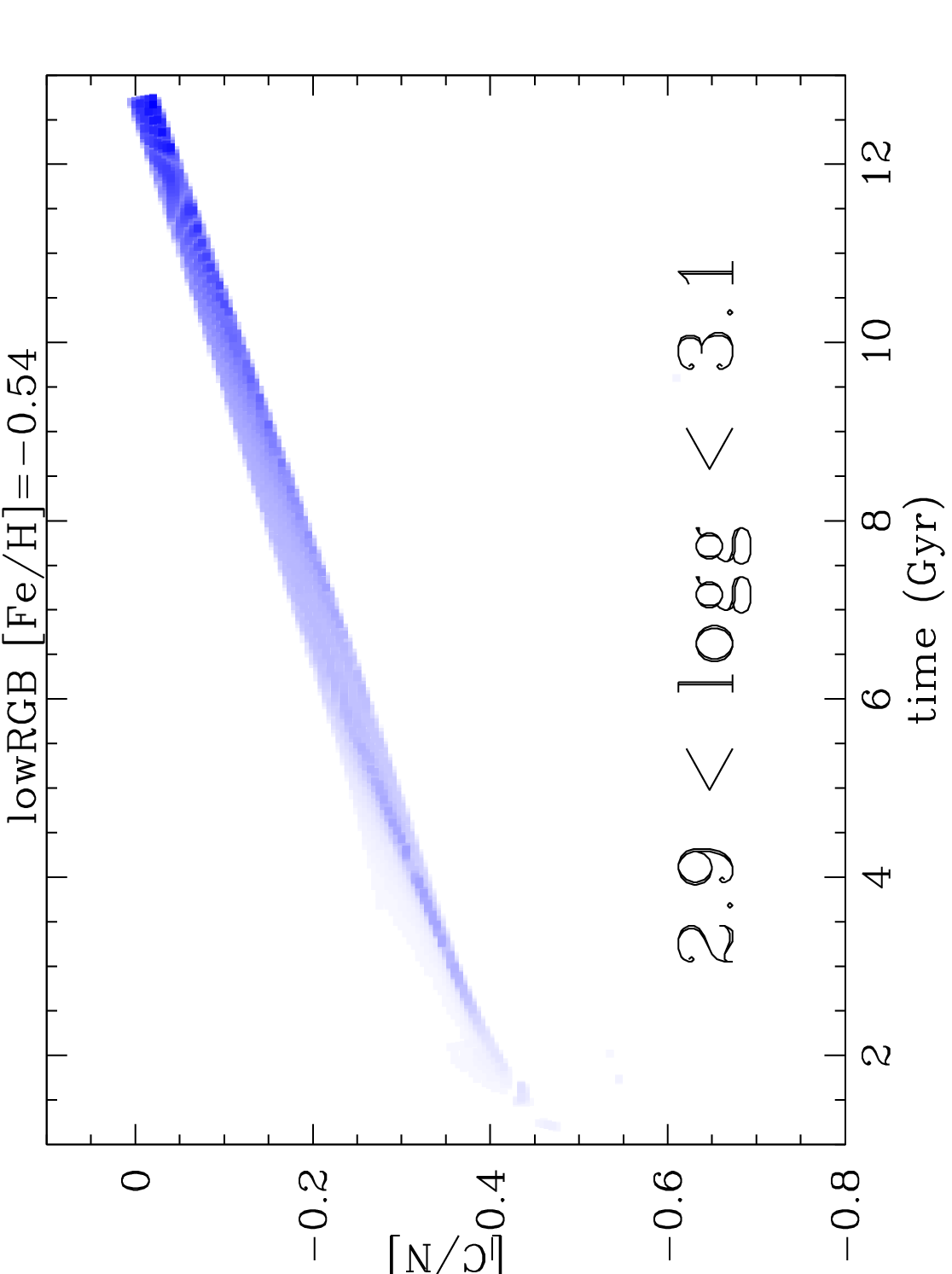}
\caption{Same as Fig~\ref{fig:CNvstime_lowRGB} but with different $\log g$ cuts. In general the trends are not affected by changes in our selection criteria, which only affects the scatter. }\label{fig:CNvstime_lowRGB_loggcuts}
\end{figure}

However, as illustrated in Fig.~\ref{fig:CNvslogg} there are discrepancies between the models and the observations. Obvious examples evident by visual inspection include the predicted vs the observed location of the Red Clump, and the inconsistency between the predicted low values for [C/N] at low $\log g$ and the observed high values.   One can expect our simulation results to be affected by such disagreements, in particular regarding our selection criteria in $\log g$. We do not know here whether the differences are a problem in the APOGEE calibration for surface gravities or whether the models are not accurate enough, and we further recall that we have offset the APOGEE [C/N] zero point to match Solar, and thus to improve agreement with the models.  We can however test the model sensitivity, and so have run the same simulation with different $\log g$ cuts. We demonstrate in Fig.~\ref{fig:CNvstime_lowRGB_loggcuts} that varying the $\log g$ selection does affect relative stellar densities, but that overall the trends remain identical, as do our conclusions.

\subsubsection{Simulation results for the lower RGB stars and discussion}
 From Fig.~\ref{fig:CNvstime_lowRGB} and using the results from our albeit preliminary stellar evolutionary simulations, we can now interpret the various trends observed for both disk populations. As seen in the simulations, the [C/N] value shows an overall decrease when the metallicity increases, mostly driven by the decrease of the initial [C/N]. This explains why the general trends for the thin and thick disk stars in Fig.~\ref{fig:CNvstime_lowRGB} are negative. The fairly small range in [C/N] for the thick disk stars is significant.  In fact, our simulation indicates that low metallicity thick disk stars have approximately the same age as do the most metal-rich thick disk stars.  Indeed, this is a further proof that the thick disk had a fast star formation and self-enrichment history, consistent with the long-standing deduction from the enhanced [$\alpha$/Fe] values. However, we observe for the thin disk stars a steeper trend in [C/N] as a function of [Fe/H] than for the thick disk. Interpreting the [C/N] values for the thin disk in terms of our simulations, this clearly indicates that there is an age difference of $\approx$6Gyr between the metal-poor thin disk stars and the solar abundance thin disk stars. This implies that the star formation history has been much slower in the thin disk than in the thick disk. 

Interestingly, the thin and thick disk stellar trends seem to converge in their value of [C/N] at [Fe/H]$\approx$-0.7. This would imply that both populations have similar ages at this metallicity, which corresponds to an age of 10 to 12 Gyr, confirming the suspicions of \citet{Haywood2013} and \citet{Bensby2014}.   Although the absolute time values are quite different, and thus shed light on uncertainties affecting models and observations, our conclusions are qualitatively comparable with the work of \citet{Haywood2013}. Without detailed modelling of the sample selection function, spatial distribution, radial gradients, and so on, we cannot in this study deduce more specific age information.

 Although our analysis allows us to draw robust, albeit qualitative, conclusions when discussing the relative age and evolution of the thin and thick disk, we note that the absolute age derived from Fig.~\ref{fig:CNvstime_lowRGB}, for the thin and thick disk respectively, agrees  satisfactorily with the standard expectations (and in particular the solar age). Note that we make important assumptions regarding the cross-comparison of abundance ratios in the thin and thick disk stars i) the absolute value of [C/N] in the APOGEE data analysis pipeline is not biased by the $\alpha$ abundance; ii) the first dredge-up occurring in the two disk stellar populations is identical, meaning that it is insensitive to the special properties of thick disk stars, particularly their high [$\alpha$/Fe] abundances. Although any $\alpha$-abundance dependence on stellar dredge-up does not appear as clearly in the current data as does the evidence for non-canonical extra mixing we discuss in the following section, this question requires further study.

\subsection{The upper RGB stars: constraints on non-canonical extra mixing?}
\subsubsection{Observations}
\begin{figure}
\includegraphics[width=6cm,angle=-90]{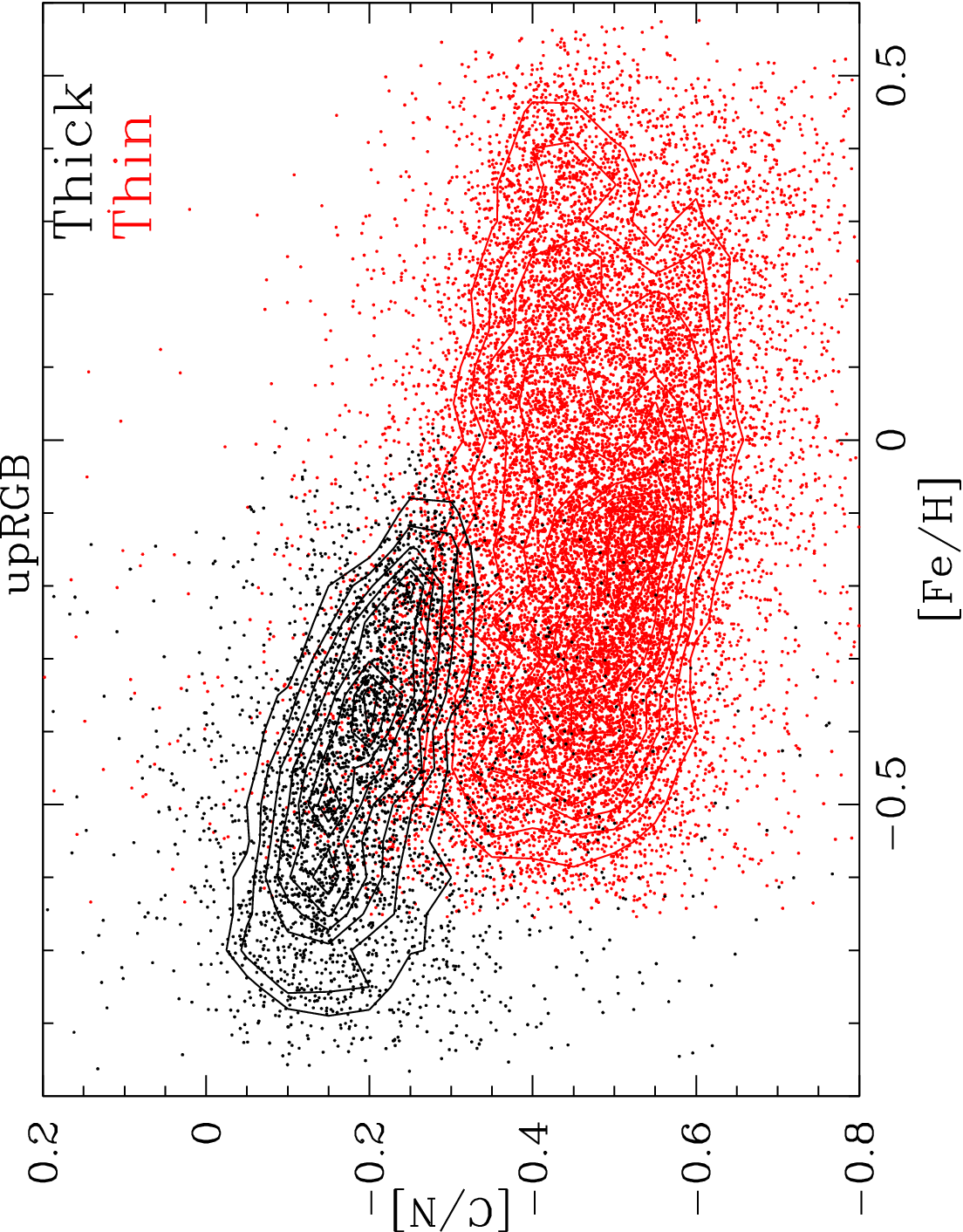}
   \caption{C/N ratio as a function of metallicity for upper RGB stars, with thick and thin disk stars classified from Fig.~\ref{fig:alphafe} and with N abundances corrected by +0.2.  While thin disk stars show a low and almost flat level of [C/N], the thick disk stars have a high [C/N] ratio. }\label{fig:CNvsFe_upRGB}.
\end{figure}

We now consider the upper RGB stars. The [C/N] vs [Fe/H] distribution for upper RGB stars is shown in Fig.\ref{fig:alphafe}, while the corresponding distribution of [C/N] with [Fe/H] is shown in Fig.~\ref{fig:CNvsFe_upRGB}. It is interesting to compare these distributions with those seen on the lower RGB, and with the model expectations shown in Fig.~\ref{fig:CNvslogg}.

The differences between Fig.\ref{fig:CNvsFe_lowRGB} and Fig.\ref{fig:CNvsFe_upRGB} are a signature of stellar evolution. As discussed above, after the first dredge-up occurs, as a star evolves further up the RGB, its C/N ratio is further depleted by non-canonical extra mixing. The favoured physical mechanism responsible for this decrease is thermohaline mixing. Although this physical mechanism is a well-known physical process (also known as salt finger effect and is observed in the oceans), \citet{Charbonnel2007} introduced it in their models and found that it is expected to set in after the first dredge-up when the star reaches the RGB luminosity bump, when the hydrogen-burning shell encounters the chemical discontinuity created inside the star by the convective envelope at its maximum extent during the first dredge-up. This supplementary mixing allows more material from H burning to be mixed and brought to the surface. More support for the idea that the driving mechanism is related to the RGB bump has been illustrated by the observation of Li abundance change along the RGB branch in a globular cluster \citep{Lind2009}. Hence, because of this mixing, it appears natural that the C/N ratios are globally lower in Fig.~\ref{fig:CNvsFe_upRGB} than in Fig.\ref{fig:CNvsFe_lowRGB}. However a closer inspection reveals that thin disk stars have a constant C/N ratio while thick disk stars show higher values and a trend. 

\subsubsection{Simulation of the [C/N] ratio in upper RGB stars}
\begin{figure*}
\includegraphics[width=6cm,angle=-90]{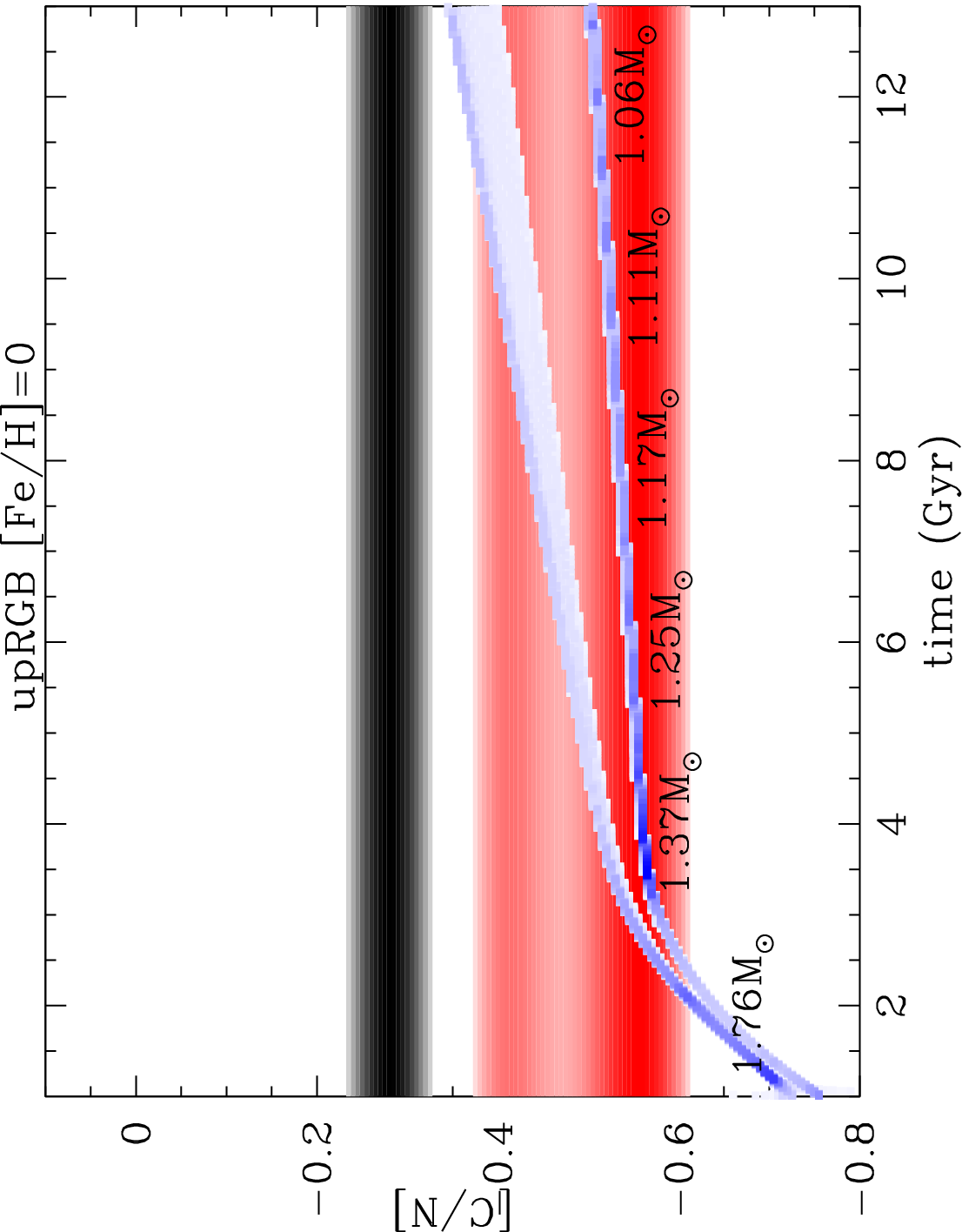}
\includegraphics[width=6cm,angle=-90]{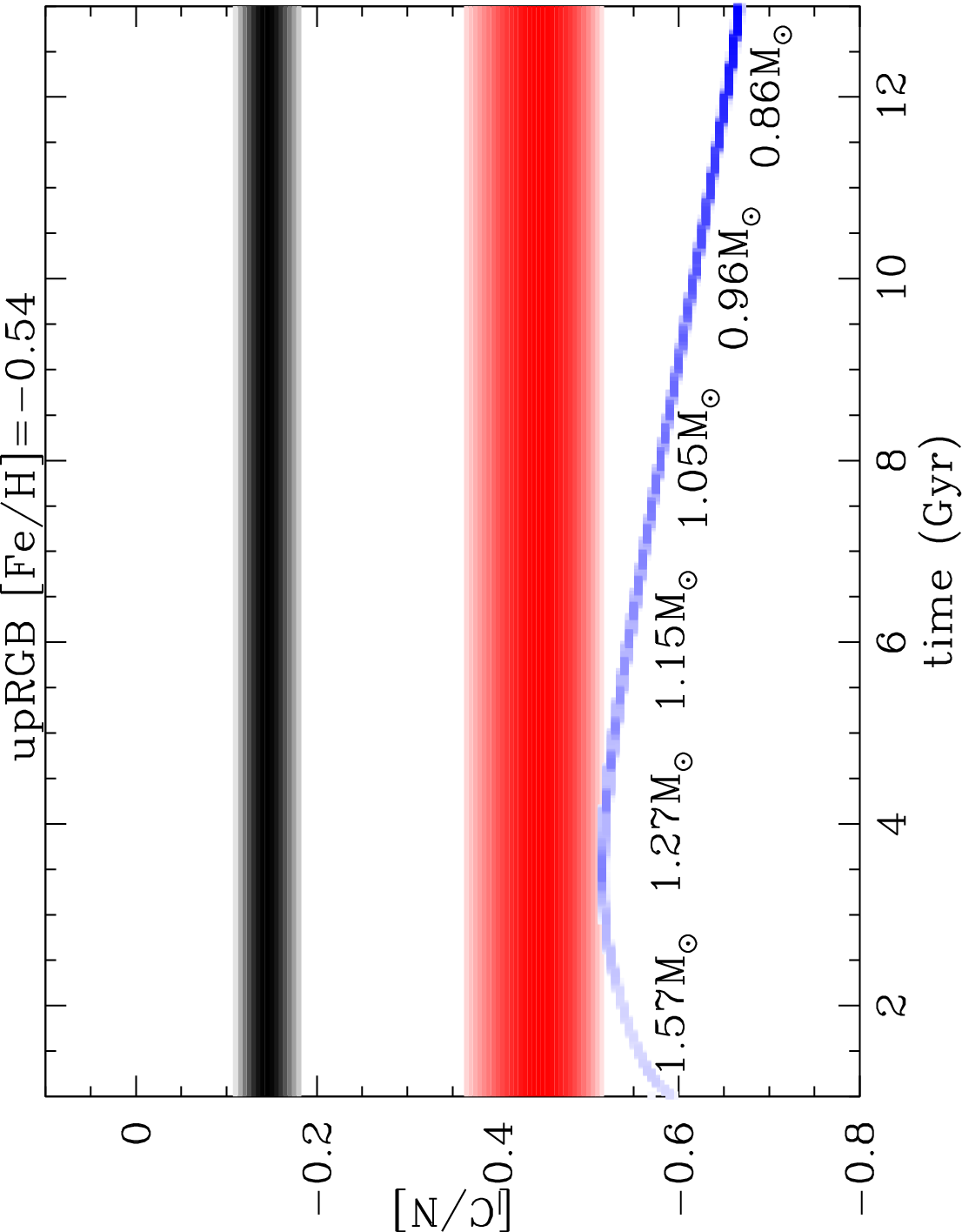}
  \caption{Simulation of evolution of [C/N] as a function of time using stellar grid from \citet{Lagarde2012}. The red and the grey area represent the values as observed in Fig.~\ref{fig:CNvsFe_upRGB}  in respectively the thin and thick disk. While the values of C/N ratios in the thin disk approximately match the expectation of stellar evolution models, it appears very difficult to reproduce the C/N ratios in the thick disk.}\label{fig:CNvstime_upRGB}
\end{figure*}

As for the first dredge-up it is expected that there is a dependence of the thermohaline mixing on the mass and metallicity of the star. To disentangle this degeneracy, we simulate the expected C/N using the same modelling approach as we implemented earlier for the lower RGB stars (Fig.~\ref{fig:CNvstime_upRGB}).  The C/N ratio from the (corrected) observations broadly matches the model expectation for the thin disk stars, although the distribution of values tends to values of [C/N] more positive than [C/N]=$-0.4$, which is not consistent with the range predicted by the models. For thick disk stars, where the observed distribution function has $\rm -0.3 < [C/N] < -0.1$, none of the models reproduces the observations.

\begin{figure}
\includegraphics[width=3cm,angle=-90]{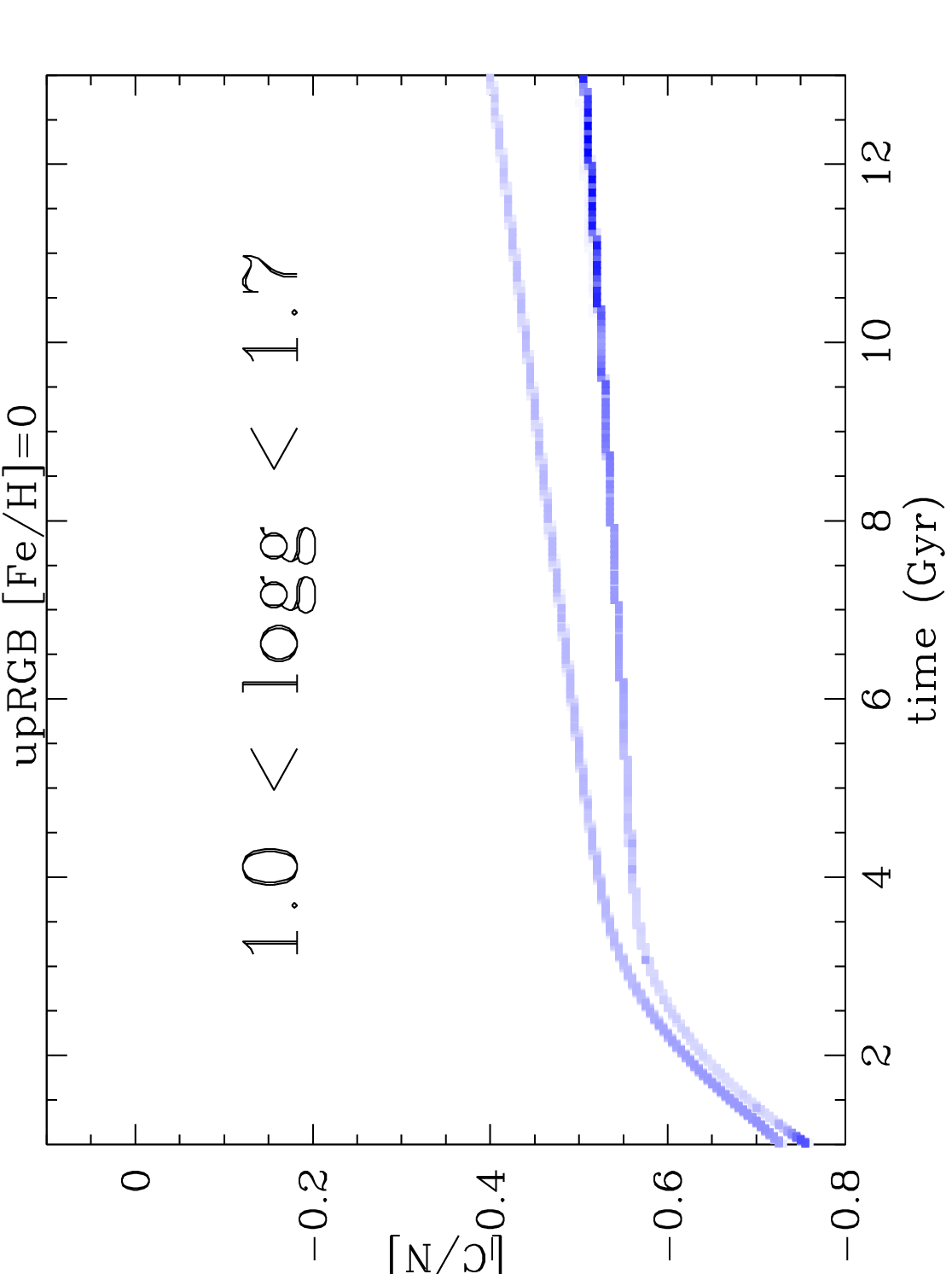}
\includegraphics[width=3cm,angle=-90]{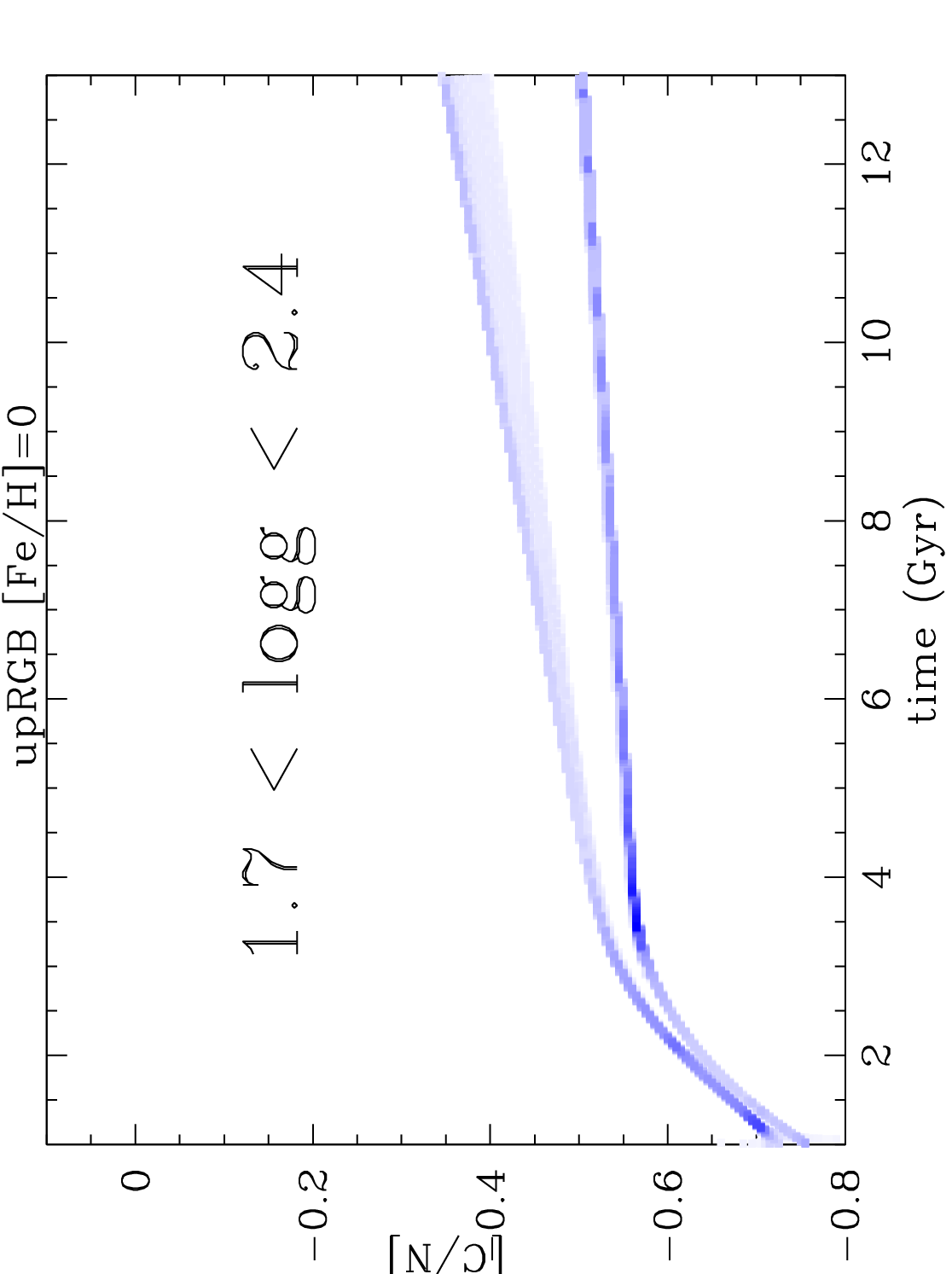}
\includegraphics[width=3cm,angle=-90]{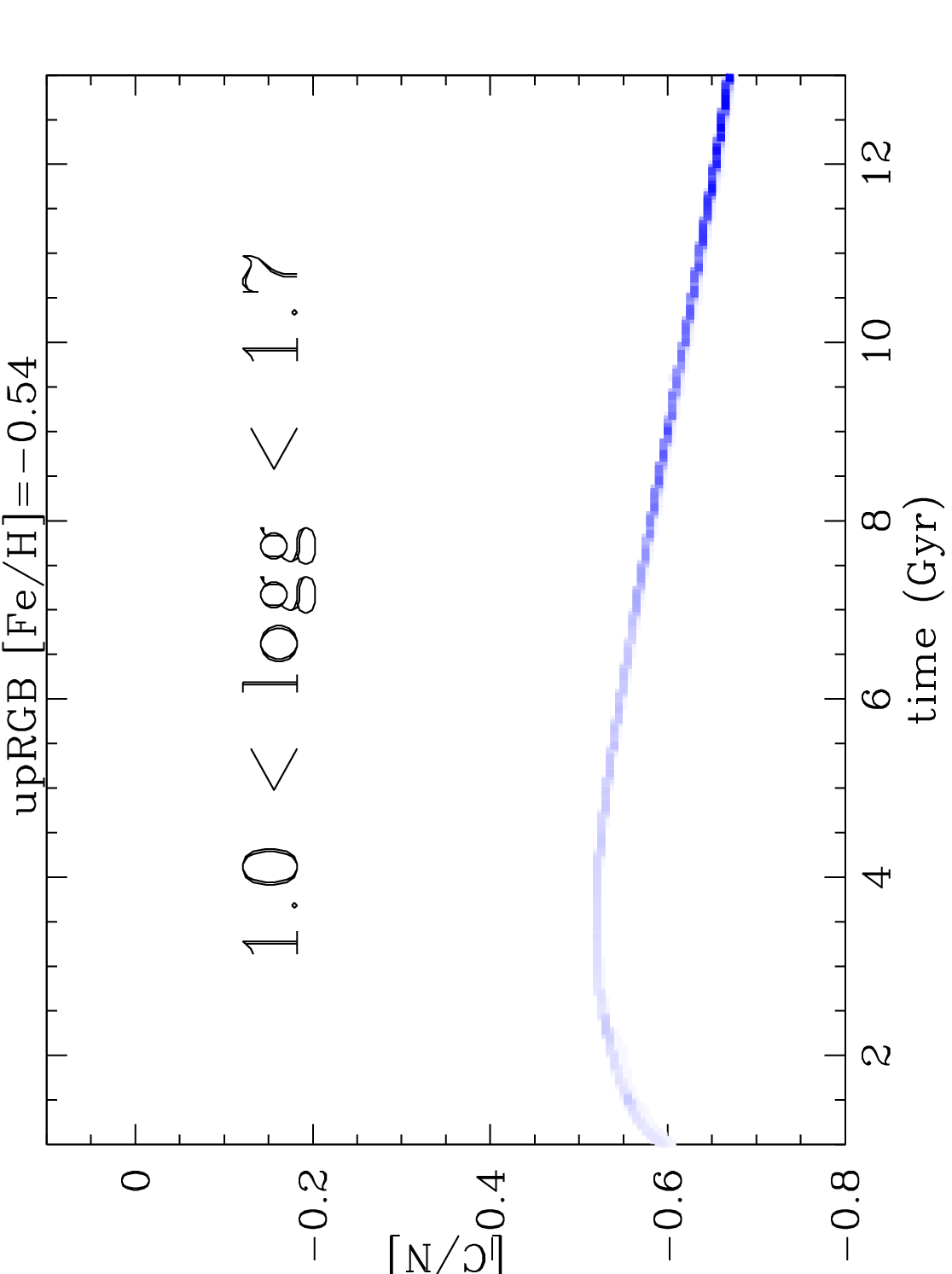}
\includegraphics[width=3cm,angle=-90]{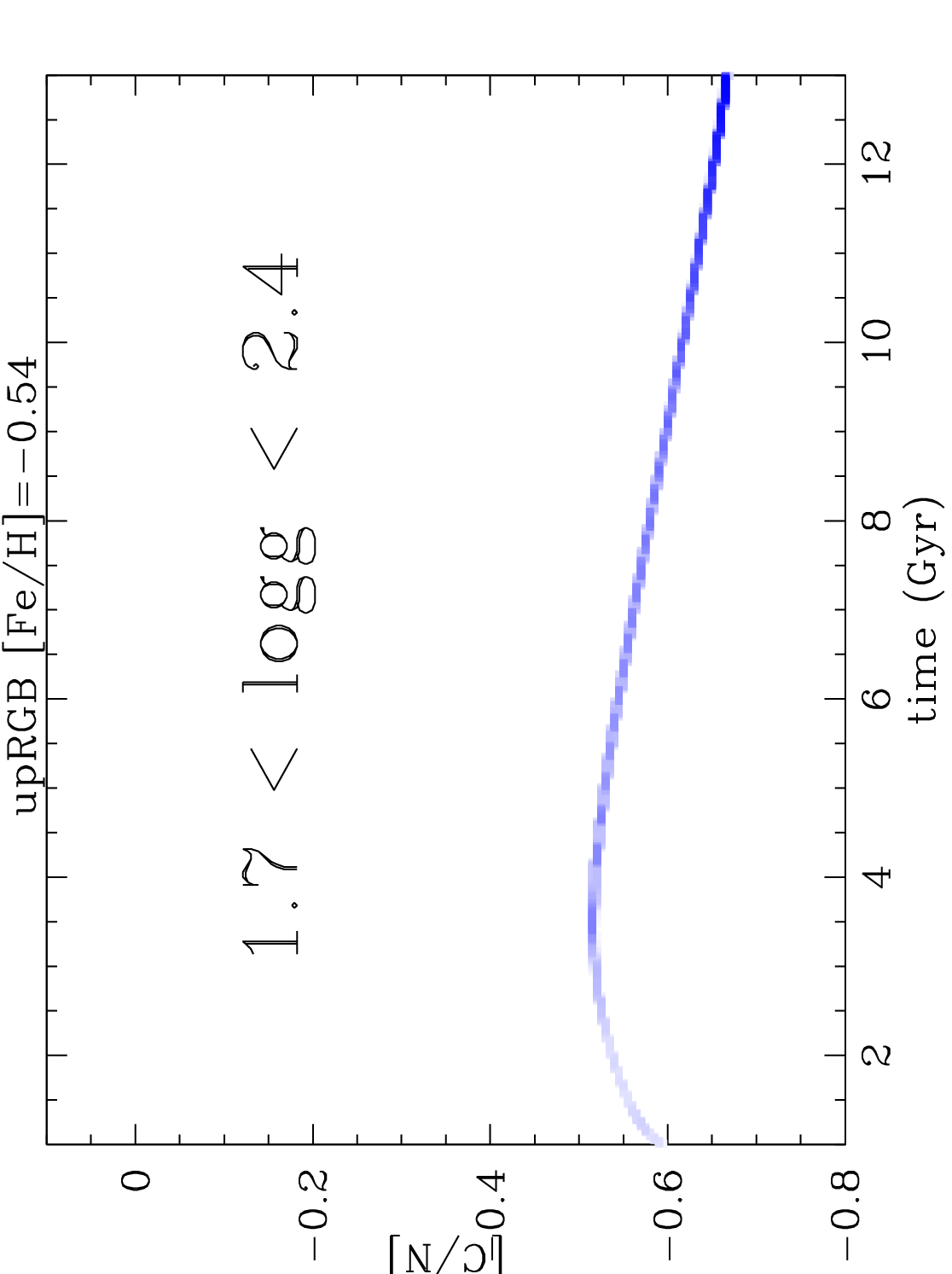}
  \caption{Same as Fig~\ref{fig:CNvstime_upRGB} but with different $\log g$ cuts. In general the trends are not affected by changes in our selection criteria, which only affects the scatter. }\label{fig:CNvstime_upRGB_loggcuts}
\end{figure}

Since [C/N] evolution on the upper RGB is very sensitive to evolutionary state, one may be concerned by our selection criteria and the disagreement between the $\log g$ provided by the models and the observed values. This could be more crucial as our definition of  upper RGB stars also encompasses several distinct stage of evolution (upper RGB branch, possibly some contamination by clump stars and even some early AGB stars). Once again to test the robustness of our conclusions, we run the simulation with different choices in $\log g$ ( Fig.~\ref{fig:CNvstime_upRGB_loggcuts}). It appears that, as for the lower RGB models, while $\log g$ selection affects the star densities in the diagram, the trends remain unchanged.   

\subsubsection{Simulation results for the upper RGB stars and discussion}
\begin{figure}
\includegraphics[width=6cm,angle=-90]{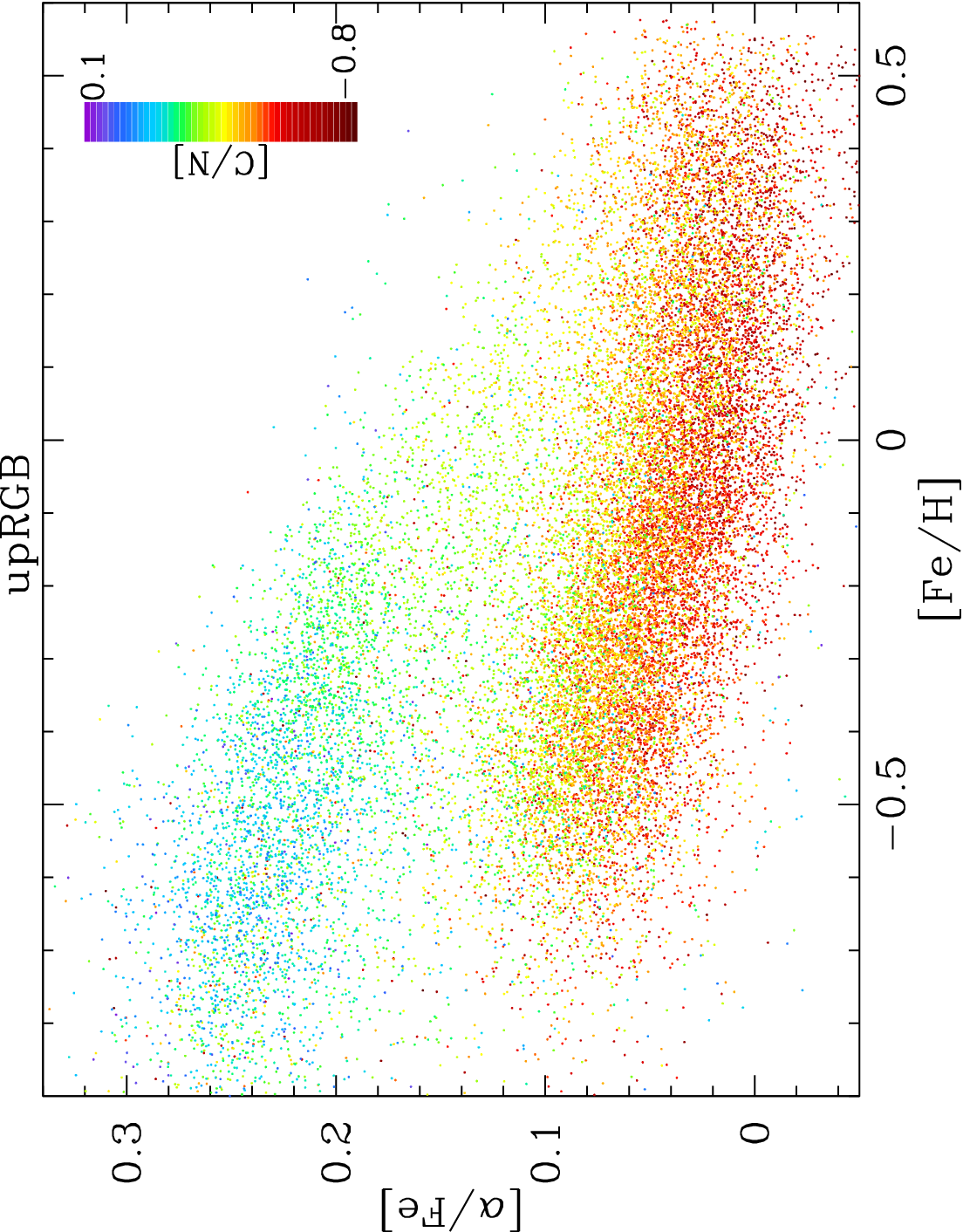}
  \caption{$\rm[\alpha/Fe]$ as a function of metallicity for upper RGB stars. The C/N ratios are color-coded. We can observe a C/N gradient proportional to $\alpha$ enhancement for the thin disks stars.}\label{fig:alphavsFe_CN_rainbow_upRGB}
\end{figure}

In detail, when comparing the lower and upper RGB thick disk stars in Fig.\ref{fig:CNvsFe_lowRGB} and Fig.\ref{fig:CNvsFe_upRGB}, any decrease in C/N ratio along the RGB  appears relatively weak, testifying that the non-canonical extra mixing has been significantly reduced. This is the first time that such an effect of reduced non-canonical extra mixing on the RGB branch has been observed in a large population.  In the literature the only studied mechanism that could inhibit non-canonical extra mixing is a magnetic field, but this is expected to affect only decendents of Ap stars \citep{Charbonnel2007_inhib}, which represent only 5\% of the stellar population. Because the thin and thick disk star samples studied here have similar properties except $\alpha$ element composition, this suggests that this may be the important factor affecting non-canonical extra mixing. An extra hint towards this hypothesis is provided by Fig.~\ref{fig:alphavsFe_CN_rainbow_upRGB} where a C/N gradient proportional to the  $\alpha$ enhancement is observed in the thin disk stars. On the other hand, at lower metallicities and even higher $\alpha$-enhancement, it has been well-established since the work of \citet{Spite2005} that canonical extra mixing is very efficient ($[C/N] \approx -1.0$) in $\alpha$ enhanced halo stars.{\bf If thermohaline mixing is indeed the physical mechanism \citep{Charbonnel2007,Lagarde2012}, more studies may be required to be able to evaluate its efficiency between very low metallicity halo stars and relatively more metal-rich thick disk stars, as well as the impact of various composition environments such as the one between thin and thick disks.   } 

\section{Putting the sample together}
\subsection{C+N vs O}
\begin{figure}
\includegraphics[width=6cm,angle=-90]{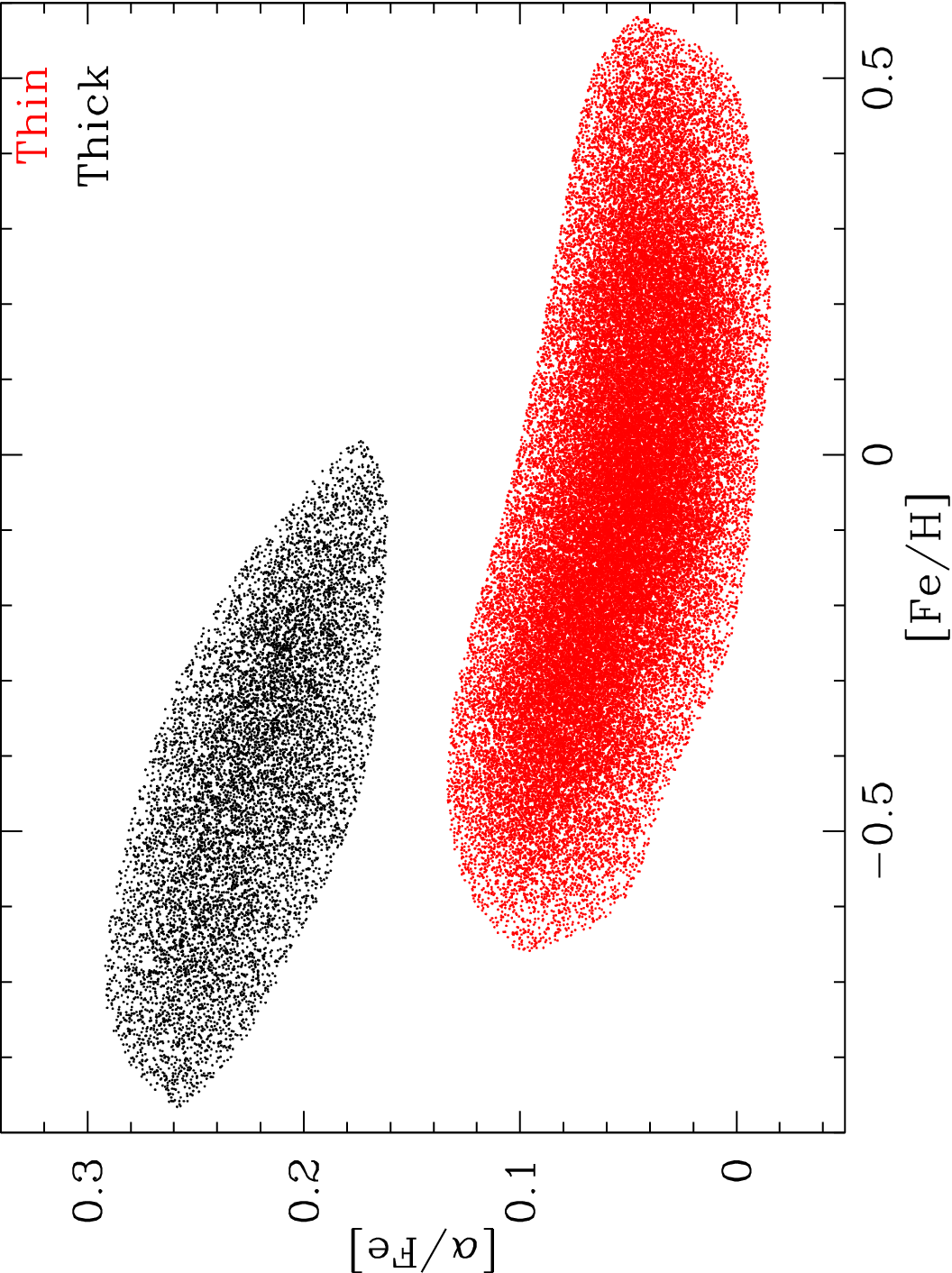}
\includegraphics[width=6cm,angle=-90]{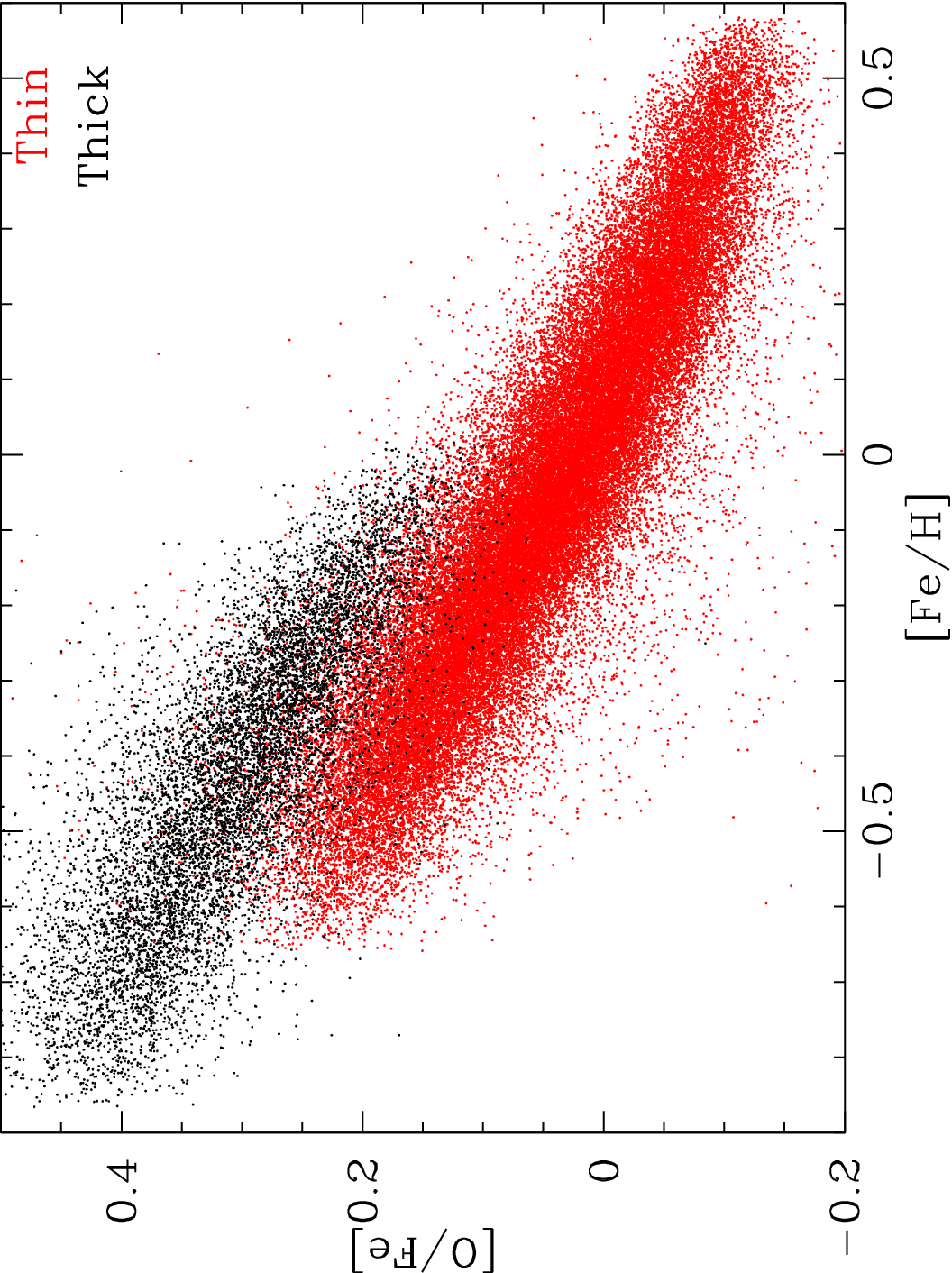}
\includegraphics[width=6cm,angle=-90]{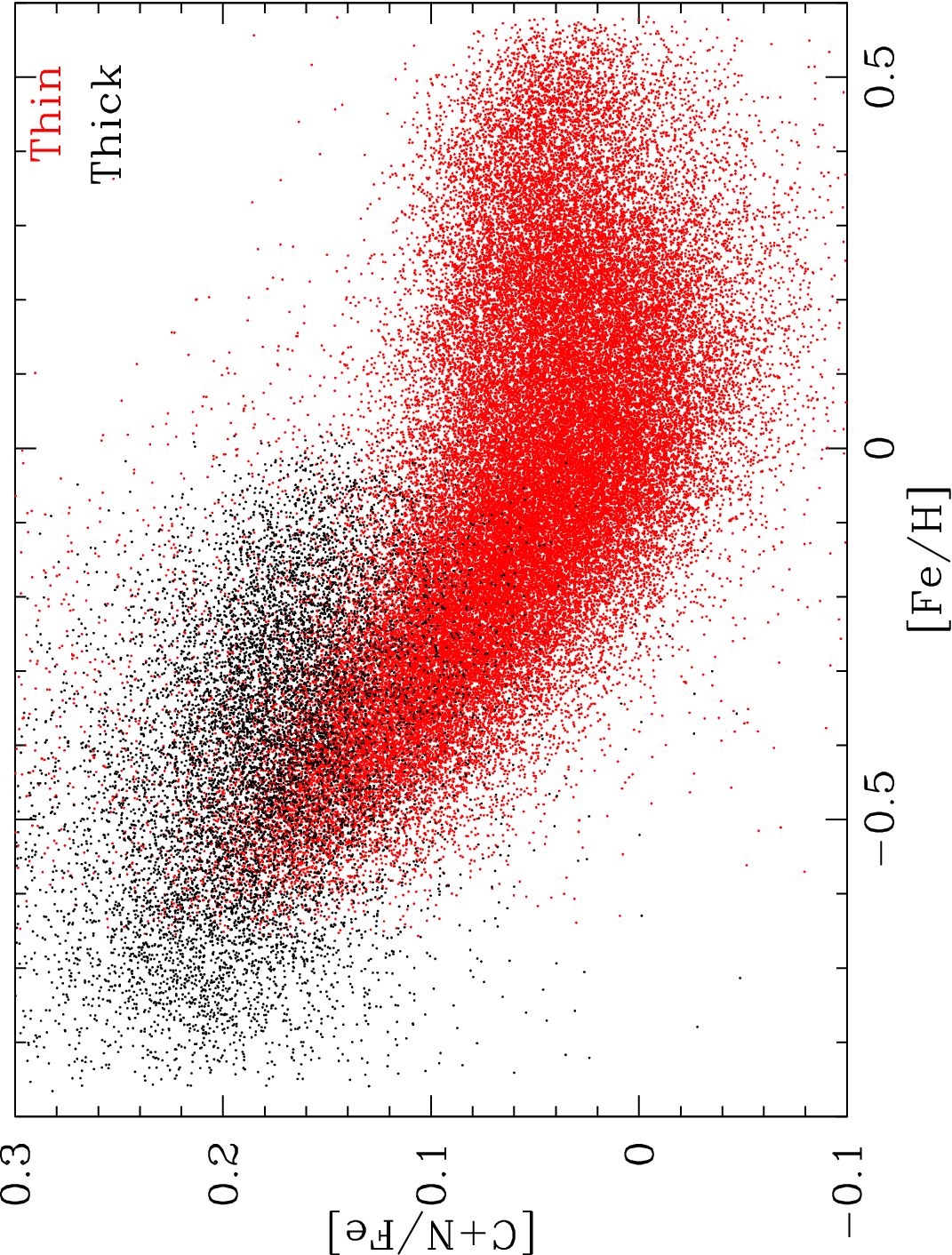}
  \caption{Element ratios as indicated for all classified thick disk (high-$\alpha$) and thin disk stars in the APOGEE RGB sample. Recall that we have offset the [N/H] zero point to match solar values.}\label{fig:C+NFeandOFevsFe}
\end{figure}

While a star's C and N abundances are affected by the star's evolution, the sum abundance, C+N, should be conserved regardless of the evolutionary status for a low-mass star, and is representative of its formation conditions. Using this property, we will in this section be able to constrain the formation of the Galactic population by now inferring the properties of the progenitors of the C+N.  We summarise the distributions of various elements and element ratios for the APOGEE giant sample in Fig.~\ref{fig:C+NFeandOFevsFe}. We note without further discussion that the abundance scale of oxygen, through [O/Fe], is low compared to other recent studies, such as \citet{deLis2015}. Clearly our conclusions here are sensitive to further studies of calibrations.

One can interpret the various trends in Fig.~\ref{fig:C+NFeandOFevsFe} in the following way: [$\alpha$/Fe] represents the contribution ratio of core collapse supernovae versus type Ia supernovae;  [C+N/Fe] represents the enrichment  of low and intermediate mass star AGB contribution versus massive star contribution.   The lifetimes of low mass AGB stars to release their nucleosynthetic products is larger than that for SNIa, which are larger than for SN type II.  

However, Fig.~\ref{fig:C+NFeandOFevsFe} shows that thick disk stars have formed with higher initial abundances of C+N than did thin disk stars, with this high abundance being indicative of the early stages of  Galaxy evolution.  One can also see that, while the C+N enhancement in the thick disk stars  decreases only slowly with increasing [Fe/H], it decreases much faster with [Fe/H] in the thin disk. This is explained if the thick disk formed on a shorter time scale than did the thin disk, in the same range of [Fe/H], so that that SNIa had released more Fe in the thin disk.  Although low-mass AGB stars are believed to be the major sources for C and N in the discs, we can still observe a decrease in [C+N/Fe], which indicates a larger contribution of Fe from SNIa. However, the slope of [C+N/Fe] vs [Fe/H] is lower than for [O/Fe], which means that the C+N/Fe decrease is partially compensated by the contribution of low and intermediate mass AGB stars. \citet{Bensby2005} measured [Ba/Eu] and found an early s-process contribution, which confirms that AGB stars contribute to the C+N enrichment of the thin disk early on. Moreover, we observe a "knee" in C+N at [Fe/H]$\approx$0 in the thin disk. If this additional increase in C+N is related to AGB stars, then this may mean that there is a significant change in the production of carbon for AGB stars above solar metallicity.

The abundance of [C+N] is similar for both thin and thick disk stars at metallicity $\sim -0.7$.  We have shown earlier another correspondence in [C/N] at a similar metallicity, indicating that thin and thick disk had similar ages at this point. Considering also the observations of \citet{Bensby2014}, suggesting a lower limit for thin disk metallicity at $-0.7$, and the discovery of some kinematically selected thin disk alpha-rich stars at [Fe/H]$\approx -0.7$ by \citet{Haywood2013}, we seem to seeing that, for stars now in the Solar neighbourhood, thick disk enrichment continued independently of the start of formation of the thin disk between 10 and 12Gyr ago.

\subsection{The rest of the sample}

For this study we have empirically defined two dominant stellar populations, thick and thin disks, using as definition population density. We excluded from consideration, until now, sources outside the dominant groups. These apparent outliers will be a mix of errors, but also may represent both expected outliers, such as binary mass transfer stars,  and potentially interesting sub-populations.    These stars are identified in the element ratio diagrams in Fig~\ref{fig:C+NFeandOFevsFe_extrapop}. 

\subsubsection{The Intermediate $\rm \alpha$-enhancement  population}

\begin{figure}
\includegraphics[width=6cm,angle=-90]{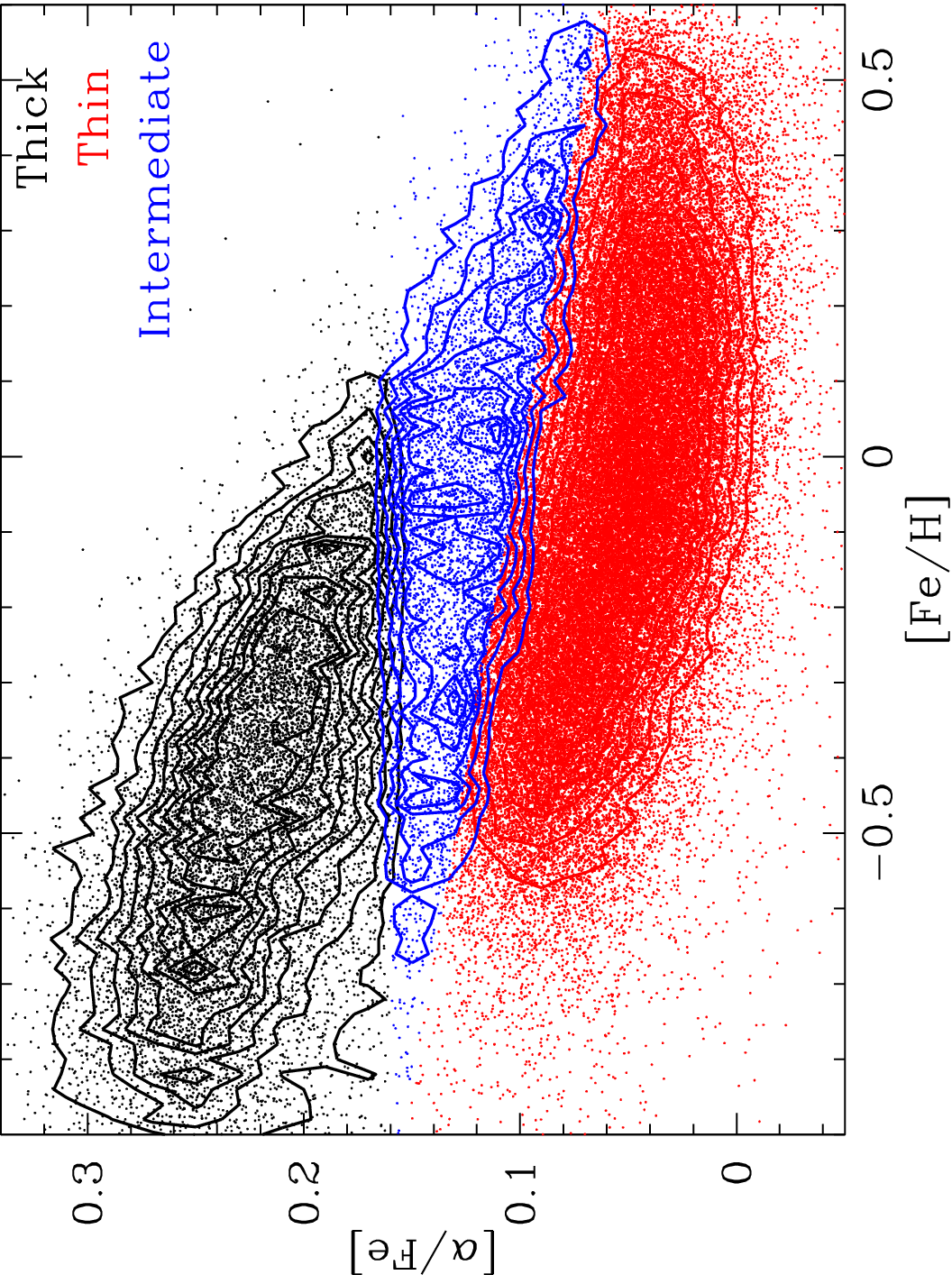}
\includegraphics[width=6cm,angle=-90]{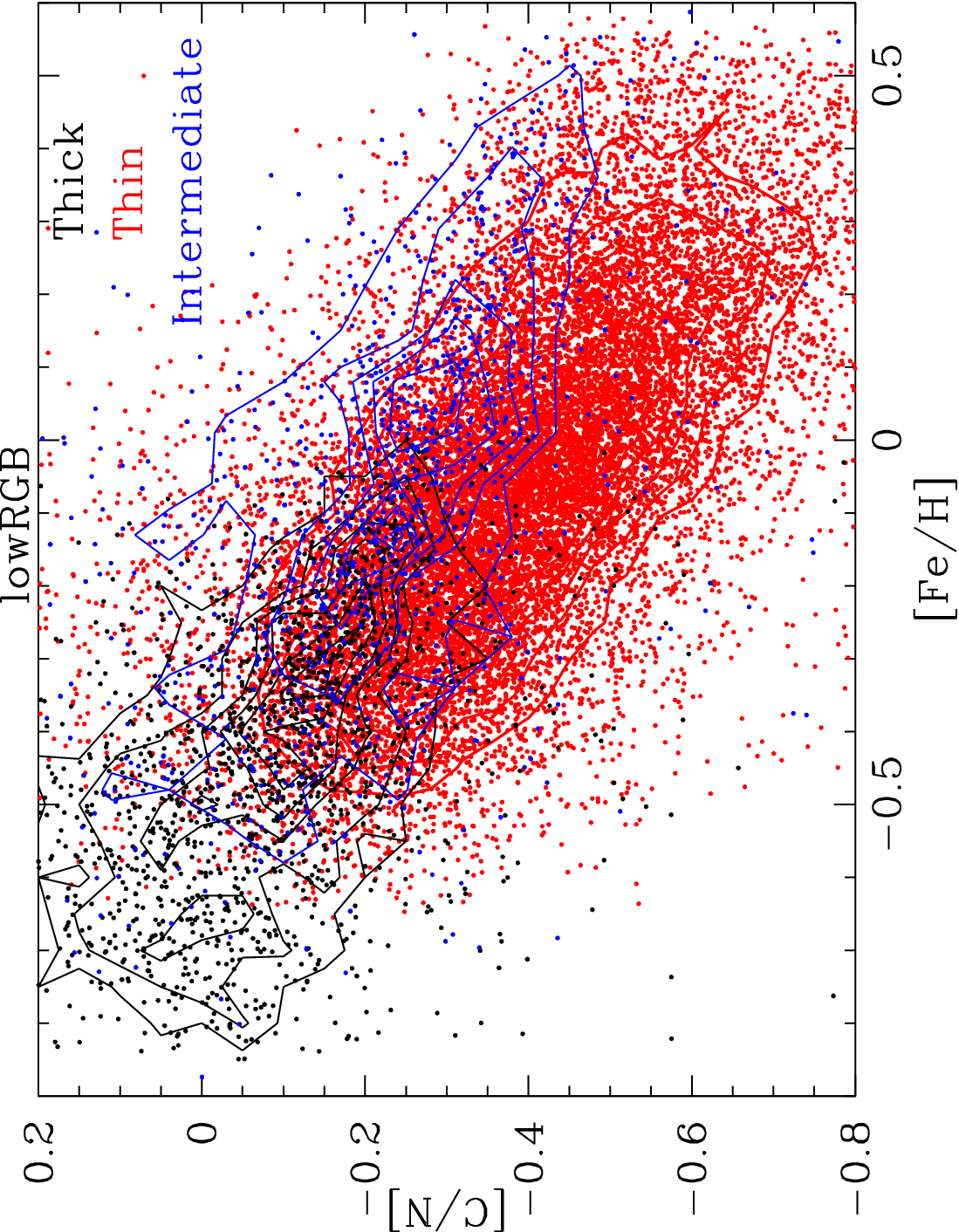}
\includegraphics[width=6cm,angle=-90]{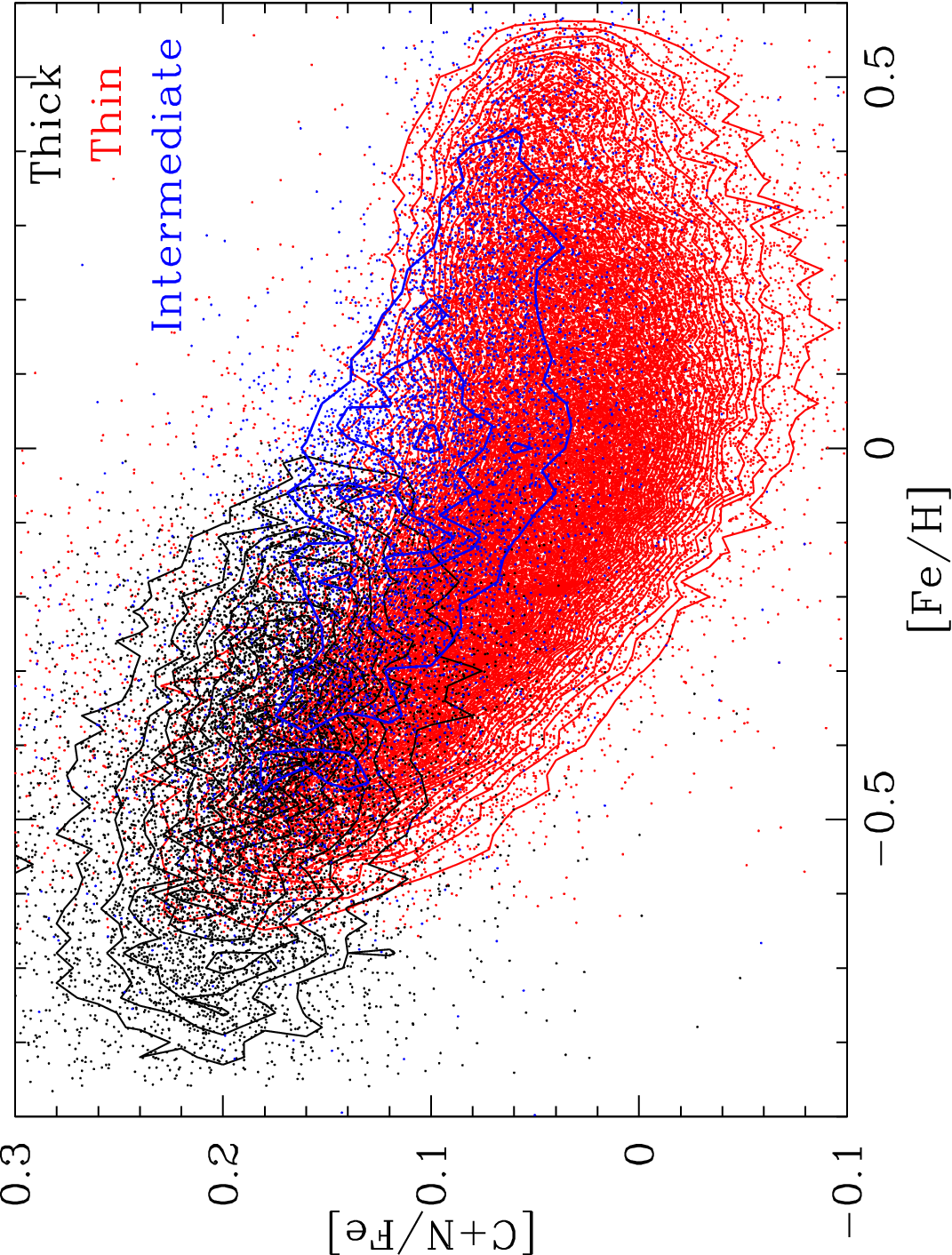}
\caption{In the top panel we define the sample of stars which lie outside the two main groups defining thick disk and thin disk. The [C/N] (middle) and [C+N] (lower) distributions of these stars are presented in the other panels (including  a zero point offset in N abundances).}\label{fig:C+NFeandOFevsFe_extrapop}
\end{figure}
Fig.~\ref{fig:C+NFeandOFevsFe_extrapop} shows the element ratio distributions of the stars which lie outside the dominant population groups. While the density of these stars is low, the majority do form a smooth extension of the trends seen in the main groups.  Further study of measurement errors and scatter are required before a detailed discussion is feasible. Among topics of future interest are to determine if the data are able to make any useful statement as to whether  the intermediate$- \alpha$ population at solar metallicity is a younger extension  of a thick-disk like population, or is an older metal-poor thin disk (migrated?) population ( \citet{Adibekyan2011, Bensby2014}), or something else.

\subsubsection{Super-solar population}
Potentially interesting sub-structure is evident at super solar metallicity in Fig.~\ref{fig:CNvsFe_upRGB}, suggesting that many stars with [Fe/H=+0.25  have [C/N]=$-0.45$, an unexpectedly high value. While this super solar population contains stars with intermediate C/N ratios, it does not show any particular $\alpha$ enhancement compared to the thin disk population (Fig.~\ref{fig:CNvsFe_upRGB}). Note also that this population does not appear as clearly in lower RGB stars, as indicated in Fig.~\ref{fig:CNvsFe_lowRGB}. This may suggest that we detect the effect of metallicity on non-canonical extra mixing on the upper RGB, rather than a true distinct population of super-metal-rich old stars. In the latter case, these stars may be related to the recent finding of \citet{Kordopatis2014}, revealing a local population likely to have radially migrated from inner regions of the Galaxy.  This group may also be related to the $\alpha -$enhanced population as seen in figure 9 of \citet{Nidever2014}. However, more theoretical study of the impact of $\alpha -$abundances and metallicity on the dredge up and on non-canonical mixing processes is needed before drawing any firm conclusions.

\section{Conclusions}

Defining the Galactic thick disk is an interesting challenge, with several combinations of kinematics and chemistry in use. We show here that there is a clear bimodality in the distribution function of stars in the [$\alpha$/Fe] vs [Fe/H] space. Further, we show that this bimodality is very highly correlated with stellar [C/N] ratio in red giant stars.  The [C/N] ratio of an RGB star is determined by its evolutionary state, and in particular is a function of stellar mass, and hence age. Our observed correlation of properties allows us to determine the relative age distribution functions of the thick and the thin disk, showing that the thick disk red giants are indeed systematically older than the thin disk red giants at the same value of [Fe/H]. Thus the thick disk is older than the thin disk. The thick disk also has higher values of [$\alpha$/Fe] ratios, implying also that it formed more rapidly than did the metal-poor stars of the thin disk.

C/N ratio is an excellent and alternative probe for constraining the relative masses of giant stars and hence their relative ages without the use of isochrones. As a quantitative comparison, determining an age of RGB branch stars from isochrone fitting with a precision of 1Gyr requires an accuracy in $\rm T_{eff}$ better than 50K while it requires an accuracy in [C/N] of 0.05. 
It is very tempting to assign masses and ages to the whole sample using the C/N ratios. However this exercise would first require us to prove that the absolute measurement of the C/N ratios in the RGB of the present sample is better that the actual excellent precision - and we emphasise that we see an offset of approximately 0.2dex in the zero point of [N/Fe] for solar abundance stars in the current APOGEE dataset. 

We also stress that although there are several independent studies to derive stellar population ages (e.g. \citet{Haywood2013}), all attempts rely on stellar evolution models. Unfortunately, despite the huge effort currently devoted, we show that the models still suffer from some caveats, perhaps even for long-studied effects such as the first dredge-up. In the  particular case of red giant branch stars, we show that more detailed models of the first dredge-up and of canonical extra mixing is required, specifically their sensitivity to the initial C, N and $\alpha -$element composition.

\section*{Acknowledgments}
We are very grateful to P. Jofre, A. Casey and C. Tout for fruitful discussion.
This work was partly supported by the European Union FP7 programme through 
ERC grant number 320360.

\bibliographystyle{mn2e}
\bibliography{CNdisk}

\begin{thebibliography}{}

\bibitem[\protect\citeauthoryear{{Adibekyan}, {Santos}, {Sousa} \&
  {Israelian}}{{Adibekyan} et~al.}{2011}]{Adibekyan2011}
{Adibekyan} V.~Z.,  {Santos} N.~C.,  {Sousa} S.~G.,    {Israelian} G.,  2011,
  \aap, 535, L11

\bibitem[\protect\citeauthoryear{{Anders}}{{Anders}}{2014}]{Anders2014}
{Anders} F. e.~a.,  2014, \aap, 564, A115

\bibitem[\protect\citeauthoryear{{Bensby}, {Feltzing}, {Lundstr{\"o}m} \&
  {Ilyin}}{{Bensby} et~al.}{2005}]{Bensby2005}
{Bensby} T.,  {Feltzing} S.,  {Lundstr{\"o}m} I.,    {Ilyin} I.,  2005, \aap,
  433, 185

\bibitem[\protect\citeauthoryear{{Bensby}, {Feltzing} \& {Oey}}{{Bensby}
  et~al.}{2014}]{Bensby2014}
{Bensby} T.,  {Feltzing} S.,    {Oey} M.~S.,  2014, \aap, 562, A71

\bibitem[\protect\citeauthoryear{{Bertran de Lis}, {Delgado Mena}, {Adibekyan},
  {Santos} \& {Sousa}}{{Bertran de Lis} et~al.}{2015}]{deLis2015}
{Bertran de Lis} S.,  {Delgado Mena} E.,  {Adibekyan} V.~Z.,  {Santos} N.~C.,
   {Sousa} S.~G.,  2015, ArXiv e-prints

\bibitem[\protect\citeauthoryear{{Bovy}, {Rix}, {Liu}, {Hogg}, {Beers} \&
  {Lee}}{{Bovy} et~al.}{2012}]{Bovy2012}
{Bovy} J.,  {Rix} H.-W.,  {Liu} C.,  {Hogg} D.~W.,  {Beers} T.~C.,    {Lee}
  Y.~S.,  2012, \apj, 753, 148

\bibitem[\protect\citeauthoryear{{Charbonnel}}{{Charbonnel}}{1994}]{Charbonnel%
1994}
{Charbonnel} C.,  1994, \aap, 282, 811

\bibitem[\protect\citeauthoryear{{Charbonnel} \& {Zahn}}{{Charbonnel} \&
  {Zahn}}{2007a}]{Charbonnel2007_inhib}
{Charbonnel} C.,  {Zahn} J.-P.,  2007a, \aap, 476, L29

\bibitem[\protect\citeauthoryear{{Charbonnel} \& {Zahn}}{{Charbonnel} \&
  {Zahn}}{2007b}]{Charbonnel2007}
{Charbonnel} C.,  {Zahn} J.-P.,  2007b, \aap, 467, L15

\bibitem[\protect\citeauthoryear{{De Silva} \& et al.}{{De Silva} \&
  et~al.}{2015}]{DeSilva2015}
{De Silva} G.~M.,  et al. 2015, ArXiv e-prints

\bibitem[\protect\citeauthoryear{{Fuhrmann}}{{Fuhrmann}}{1998}]{Fuhrmann1998}
{Fuhrmann} K.,  1998, \aap, 338, 161

\bibitem[\protect\citeauthoryear{{Fuhrmann}}{{Fuhrmann}}{2011}]{Fuhrmann2011}
{Fuhrmann} K.,  2011, \mnras, 414, 2893

\bibitem[\protect\citeauthoryear{{Gilmore}, {Randich}, {Asplund}, {Binney},
  {Bonifacio}, {Drew}, {Feltzing}, {Ferguson}, {Jeffries}, {Micela} \& et
  al.}{{Gilmore} et~al.}{2012}]{Gilmore2012}
{Gilmore} G.,  {Randich} S.,  {Asplund} M.,  {Binney} J.,  {Bonifacio} P.,
  {Drew} J.,  {Feltzing} S.,  {Ferguson} A.,  {Jeffries} R.,  {Micela} G.,
  et al. 2012, The Messenger, 147, 25

\bibitem[\protect\citeauthoryear{{Gilmore} \& {Reid}}{{Gilmore} \&
  {Reid}}{1983}]{Gilmore1983}
{Gilmore} G.,  {Reid} N.,  1983, \mnras, 202, 1025

\bibitem[\protect\citeauthoryear{{Gratton}, {Sneden}, {Carretta} \&
  {Bragaglia}}{{Gratton} et~al.}{2000}]{Gratton2000}
{Gratton} R.~G.,  {Sneden} C.,  {Carretta} E.,    {Bragaglia} A.,  2000, \aap,
  354, 169

\bibitem[\protect\citeauthoryear{{Guiglion}, {Recio-Blanco} \& {de
  Laverny}}{{Guiglion} et~al.}{2015}]{Guiglion2015}
{Guiglion} G.,  {Recio-Blanco} A.,    {de Laverny} P.,  2015, \aap, 999, XXX

\bibitem[\protect\citeauthoryear{{Haywood}, {Di Matteo}, {Lehnert}, {Katz} \&
  {G{\'o}mez}}{{Haywood} et~al.}{2013}]{Haywood2013}
{Haywood} M.,  {Di Matteo} P.,  {Lehnert} M.~D.,  {Katz} D.,    {G{\'o}mez} A.,
   2013, \aap, 560, A109

\bibitem[\protect\citeauthoryear{{Holtzman} \& et al.}{{Holtzman} \&
  et~al.}{2015}]{Holtzman2015}
{Holtzman} J.~A.,  et al. 2015, ArXiv e-prints

\bibitem[\protect\citeauthoryear{{Iben} Jr.}{{Iben}}{1965}]{Iben1965}
{Iben} Jr. I.,  1965, \apj, 142, 1447

\bibitem[\protect\citeauthoryear{{Israelian}, {Ecuvillon}, {Rebolo},
  {Garc{\'{\i}}a-L{\'o}pez}, {Bonifacio} \& {Molaro}}{{Israelian}
  et~al.}{2004}]{Israelian2004}
{Israelian} G.,  {Ecuvillon} A.,  {Rebolo} R.,  {Garc{\'{\i}}a-L{\'o}pez} R.,
  {Bonifacio} P.,    {Molaro} P.,  2004, \aap, 421, 649

\bibitem[\protect\citeauthoryear{{Kordopatis} \& et al.}{{Kordopatis} \&
  et~al.}{2014}]{Kordopatis2014}
{Kordopatis} G.,  et al. 2014, ArXiv e-prints

\bibitem[\protect\citeauthoryear{{Lagarde}, {Decressin}, {Charbonnel},
  {Eggenberger}, {Ekstr{\"o}m} \& {Palacios}}{{Lagarde}
  et~al.}{2012}]{Lagarde2012}
{Lagarde} N.,  {Decressin} T.,  {Charbonnel} C.,  {Eggenberger} P.,
  {Ekstr{\"o}m} S.,    {Palacios} A.,  2012, \aap, 543, A108

\bibitem[\protect\citeauthoryear{{Lee}, {Beers}, {An}, {Ivezi{\'c}}, {Just},
  {Rockosi}, {Morrison}, {Johnson}, {Sch{\"o}nrich}, {Bird}, {Yanny}, {Harding}
  \& {Rocha-Pinto}}{{Lee} et~al.}{2011}]{Lee2011}
{Lee} Y.~S.,  {Beers} T.~C.,  {An} D.,  {Ivezi{\'c}} {\v Z}.,  {Just} A.,
  {Rockosi} C.~M.,  {Morrison} H.~L.,  {Johnson} J.~A.,  {Sch{\"o}nrich} R.,
  {Bird} J.,  {Yanny} B.,  {Harding} P.,    {Rocha-Pinto} H.~J.,  2011, \apj,
  738, 187

\bibitem[\protect\citeauthoryear{{Lind}, {Primas}, {Charbonnel}, {Grundahl} \&
  {Asplund}}{{Lind} et~al.}{2009}]{Lind2009}
{Lind} K.,  {Primas} F.,  {Charbonnel} C.,  {Grundahl} F.,    {Asplund} M.,
  2009, \aap, 503, 545

\bibitem[\protect\citeauthoryear{{Magic}, {Collet}, {Hayek} \&
  {Asplund}}{{Magic} et~al.}{2013}]{Magic2013}
{Magic} Z.,  {Collet} R.,  {Hayek} W.,    {Asplund} M.,  2013, \aap, 560, A8

\bibitem[\protect\citeauthoryear{{Martell}, {Smith} \& {Briley}}{{Martell}
  et~al.}{2008}]{Martell2008}
{Martell} S.~L.,  {Smith} G.~H.,    {Briley} M.~M.,  2008, \aj, 136, 2522

\bibitem[\protect\citeauthoryear{{Mikolaitis} \& et al.}{{Mikolaitis} \&
  et~al.}{2014}]{Mikolaitis2014}
{Mikolaitis} {\v S}.,  et al. 2014, \aap, 572, A33

\bibitem[\protect\citeauthoryear{{Nidever} \& et al.}{{Nidever} \&
  et~al.}{2014}]{Nidever2014}
{Nidever} D.~L.,  et al. 2014, \apj, 796, 38

\bibitem[\protect\citeauthoryear{{Nissen}, {Chen}, {Carigi}, {Schuster} \&
  {Zhao}}{{Nissen} et~al.}{2014}]{Nissen2014}
{Nissen} P.~E.,  {Chen} Y.~Q.,  {Carigi} L.,  {Schuster} W.~J.,    {Zhao} G.,
  2014, \aap, 568, A25

\bibitem[\protect\citeauthoryear{{Oswalt} \& {Gilmore}}{{Oswalt} \&
  {Gilmore}}{2013}]{Gilmore2013}
{Oswalt} T.~D.,  {Gilmore} G.,  2013, {Planets, Stars and Stellar Systems Vol.
  5}

\bibitem[\protect\citeauthoryear{{Recio-Blanco}}{{Recio-Blanco}}{2014}]{Recio-%
Blanco2014}
{Recio-Blanco} A. e.~a.,  2014, \aap, 567, A5

\bibitem[\protect\citeauthoryear{{Reddy}, {Tomkin}, {Lambert} \& {Allende
  Prieto}}{{Reddy} et~al.}{2003}]{Reddy2003}
{Reddy} B.~E.,  {Tomkin} J.,  {Lambert} D.~L.,    {Allende Prieto} C.,  2003,
  \mnras, 340, 304

\bibitem[\protect\citeauthoryear{{Rix} \& {Bovy}}{{Rix} \&
  {Bovy}}{2013}]{Rix2013}
{Rix} H.-W.,  {Bovy} J.,  2013, \aapr, 21, 61

\bibitem[\protect\citeauthoryear{{Spite}, {Cayrel}, {Plez}, {Hill}, {Spite},
  {Depagne}, {Fran{\c c}ois}, {Bonifacio}, {Barbuy}, {Beers}, {Andersen},
  {Molaro}, {Nordstr{\"o}m} \& {Primas}}{{Spite} et~al.}{2005}]{Spite2005}
{Spite} M.,  {Cayrel} R.,  {Plez} B.,  {Hill} V.,  {Spite} F.,  {Depagne} E.,
  {Fran{\c c}ois} P.,  {Bonifacio} P.,  {Barbuy} B.,  {Beers} T.,  {Andersen}
  J.,  {Molaro} P.,  {Nordstr{\"o}m} B.,    {Primas} F.,  2005, \aap, 430, 655

\bibitem[\protect\citeauthoryear{{Tsuji}}{{Tsuji}}{1973}]{Tsuji1973}
{Tsuji} T.,  1973, \aap, 23, 411

\bibitem[\protect\citeauthoryear{{Yoshii}}{{Yoshii}}{1982}]{Yoshii1982}
{Yoshii} Y.,  1982, \pasj, 34, 365

\bibitem[\protect\citeauthoryear{{Yoshii}}{{Yoshii}}{2013}]{Yoshii2013}
{Yoshii} Y.,  2013, {Star Counts and Nature of the Galactic Thick Disk}.
p.~393

\end{thebibliography}

\label{lastpage}

\end{document}